\documentclass{emulateapj}

\usepackage{amssymb}

\def\ts     {\thinspace}
\def\kms    {\ts km\ts s$^{-1}$}
\def\etal   {{\rm et\ts al.}}
\def\msol   {$M_{\odot}$}
\def\lsol   {$L_{\odot}$}

\def\aco    {{\rm CO}($J$=1$\to$0)}
\def\bco    {{\rm CO}($J$=2$\to$1)}
\def\cco    {{\rm CO}($J$=3$\to$2)}
\def\dco    {{\rm CO}($J$=4$\to$3)}
\def\eco    {{\rm CO}($J$=5$\to$4)}
\def\fco    {{\rm CO}($J$=6$\to$5)}

\def\ccn    {{\rm CN}($N$=3$\to$2)}
\def\ccch   {{\rm C$_2$H}($N$=3$\to$2)}

\def\chcn    {{\rm HCN}($J$=3$\to$2)}

\def\chnc    {{\rm HNC}($J$=3$\to$2)}

\def\chco    {{\rm HCO$^+$}($J$=3$\to$2)}

\def\ahcni    {{\rm HCN}($J$=1$\leftarrow$0)}
\def\ahcoi    {{\rm HCO$^+$}($J$=1$\leftarrow$0)}
\def\ahnci    {{\rm HNC}($J$=1$\leftarrow$0)}
\def\bhcni    {{\rm HCN}($J$=2$\leftarrow$1)}
\def\bhcoi    {{\rm HCO$^+$}($J$=2$\leftarrow$1)}
\def\bhnci    {{\rm HNC}($J$=2$\leftarrow$1)}

\slugcomment{draft version \today, accepted for publication in the Astrophysical Journal}

\shorttitle{CO in Lensed $z$$>$2 Quasars}
\shortauthors{D.~A.\ Riechers}

\begin{document}

\title{
Molecular Gas in Lensed $z>2$ Quasar Host Galaxies and \\ the Star
Formation Law for Galaxies with Luminous Active Galactic Nuclei}

\author{Dominik A. Riechers\altaffilmark{1,2}}

\altaffiltext{1}{Astronomy Department, California Institute of
  Technology, MC 249-17, 1200 East California Boulevard, Pasadena, CA, USA
  91125; dr@caltech.edu}

\altaffiltext{2}{Hubble Fellow}


\begin{abstract}

We report the detection of luminous \bco, \cco, and \dco\ emission in
the strongly lensed high-redshift quasars B1938+666 ($z$=2.059),
HE\,0230--2130 ($z$=2.166), HE\,1104--1805 ($z$=2.322), and B1359+154
($z$=3.240), using the Combined Array for Research in Millimeter-wave
Astronomy. B1938+666 was identified in a `blind' CO redshift search,
demonstrating the feasibility of such investigations with millimeter
interferometers. These galaxies are lensing-amplified by factors of
$\mu_{\rm L}$$\simeq$11--170, and thus allow us to probe molecular gas
in intrinsically fainter galaxies than currently possible without the
aid of gravitational lensing. We report lensing-corrected intrinsic CO
line luminosities of $L'_{\rm
CO}$=0.65--21$\times$10$^9$\,K\,\kms\,pc$^2$, translating to H$_2$
masses of $M({\rm H_2})$=0.52--17$\times$10$^9$\,($\alpha_{\rm
CO}$/0.8)\msol. To investigate whether or not the AGN in luminous
quasars substantially contribute to $L_{\rm FIR}$, we study the
$L'_{\rm CO}$--$L_{\rm FIR}$ relation for quasars relative to galaxies
without a luminous AGN as a function of redshift. We find no
substantial differences between submillimeter galaxies and high-$z$
quasars, but marginal evidence for an excess in $L_{\rm FIR}$ in
nearby low-$L_{\rm FIR}$ AGN galaxies. This may suggest that an AGN
contribution to $L_{\rm FIR}$ is significant in systems with
relatively low gas and dust content, but only minor in the most
far-infrared-luminous galaxies (in which $L_{\rm FIR}$ is dominated by
star formation).

\end{abstract}


\section{Introduction}

Molecular line emission from galaxies at high redshift ($z$$>$2) has
proven to be a unique tool to investigate the stellar mass buildup in
the often heavily dust-enshrouded star-forming environments of massive
galaxies at early cosmic times. Molecular gas (CO) was detected in 26
$z$$>$2 quasar host galaxies to date (see, e.g., review by Solomon \&
Vanden Bout \citeyear{sv05}), all of which are luminous in the
rest-frame far infrared (FIR; typically $L_{\rm
FIR}$$\gtrsim$10$^{13}$\,\lsol; e.g., Wang et al.\
\citeyear{wan08}). These studies typically reveal molecular gas masses
of few times 10$^{10}$\,\msol, indicating gas-rich host galaxies for
the luminous active galactic nuclei (AGN) that are observed at optical
wavelengths. The high FIR luminosities are thought to be dominated by
dust heating from young stars formed in intense starburst events
comparable to those found in submillimeter galaxies (SMGs), with star
formation rates (SFRs) exceeding 1000\,\msol\,yr$^{-1}$. These
starburst are fueled by the massive molecular gas reservoirs, which
can maintain them for $>$10$^7$\,yr (e.g., Riechers et al.\
\citeyear{rie08}). The connection between gas mass and star formation 
is reflected in the relation between CO and FIR luminosities (e.g.,
Sanders et al.\ \citeyear{san91}; Gao \& Solomon \citeyear{gs04}),
which in turn can be used to constrain the contribution of the AGN to
$L_{\rm FIR}$ in quasars (e.g., Riechers et al.\ \citeyear{rie06}).
The presence of both luminous AGN and massive starbursts make these
objects the ideal candidates to better understand the AGN-starburst
connection back to early cosmic times.

The currently most efficient way to investigate the molecular gas
properties of quasar host galaxies with somewhat less extreme star
formation events is by facilitating the flux magnification provided by
gravitational lensing. Besides the flux boost, the spatial
magnification of lensed quasars also aids in resolving these systems,
allowing us to probe the gas reservoirs at smaller physical scales.
About one third of the CO-detected $z$$>$2 quasar host galaxies are
gravitationally lensed (9 sources), with typical magnification factors
between a few and $\sim$30 (Solomon \& Vanden Bout
\citeyear{sv05}). To study the gas properties of high redshift AGN 
host galaxies in more detail, we here enhance this sample by four
galaxies with magnification factors of 10.8--173, including double,
quadruple, sextuple, and Einstein ring lens configurations.

In this paper, we report the detection of \bco, \cco, and \dco\
emission in the far-infrared-luminous, strongly lensed quasar host
galaxies of B1938+666 ($z$=2.059), HE\,0230--2130 ($z$=2.166),
HE\,1104--1805 ($z$=2.322), and B1359+154 ($z$=3.240), using the
Combined Array for Research in Millimeter-wave Astronomy (CARMA). We
use a concordance, flat $\Lambda$CDM cosmology throughout, with
$H_0$=71\,\kms\,Mpc$^{-1}$, $\Omega_{\rm M}$=0.27, and
$\Omega_{\Lambda}$=0.73 (Spergel \etal\ \citeyear{spe03},
\citeyear{spe07}).

\section{Observations}

\begin{deluxetable*}{ l c c r c c c c c c }
\tabletypesize{\scriptsize}
\tablecaption{Summary of observations. \label{t1}}
\tablehead{
target        & line & $z_{\rm opt}$   & $\nu_{\rm obs}$\tablenotemark{a} & sideband & configuration & dates               & tracks & calibrator & $t_{\rm on}$/$t_{\rm tot}$\tablenotemark{b} \\
              &      &       & [GHz]           &   &      &                     &        &            & [hr] }
\startdata    
B1938+666     & \bco & ---   & 75.3638   &     & C/D/E   & 2010 Jul 05--Oct 10 & 3/10/4 & 1849+670   & 43.6/69.9 \\
              & \cco &       & 113.0425  &     &         &                     &        &            &           \\
HE\,0230--2130 & \cco & 2.163 & 109.360 & USB & D       & 2009 Aug 01--Sep 02 & 5      & 0132--169   & 13.3/22.9 \\
              &      &       &          &     &         &                     &        & 0204--170   &           \\
HE\,1104--1805 & \cco & 2.319 & 104.155 & USB & C       & 2010 Mar 21/22      & 2      & 1127--189   & 6.6/10.6  \\
B1359+154     & \cco & 3.235 & 81.652   & LSB & D/E     & 2009 Jul 07--Aug 15 & 1/2    & 1357+193   & 5.9/10.6  \\
              & \dco &       & 108.864  & USB & D/E     & 2009 Jun 29--Jul 27 & 2/2    &            & 7.2/13.3 \\
\tableline
total          &      &       &           &     &         &                     & 31    &            & 76.6/127.3 
\vspace{-1mm}
\enddata 
\tablenotetext{a}{Tuning frequency (where applicable), corresponding to zero velocity in Figs.~\ref{f1b} and \ref{f1}.}
\tablenotetext{b}{Time on source/total.}
\end{deluxetable*}


\subsection{Targeted CO Searches}

We used CARMA to observe the \cco\ transition line ($\nu_{\rm
rest}$=345.7960\,GHz) toward the $z$$>$2 quasars HE\,0230--2130,
HE\,1104--1805, and B1359+154, and the \dco\ line ($\nu_{\rm
rest}$=461.0408\,GHz) toward B1359+154, all of which are redshifted to
the 3\,mm atmospheric window (see Tab.~\ref{t1} for redshifted
frequencies).  All targets were observed with 14 or 15 antennas
(corresponding to 91 or 105 baselines per antenna configuration) for a
total of 14\,tracks in the C, D, and E configurations between 2009
June 29 and 2010 March 22, amounting to a total observing time of
57\,hr (33\,hr on source). 

Weather conditions scaled between acceptable and excellent for
observations at 3\,mm wavelengths. The nearby quasars J0132--169,
J0204--170, J1127--189, and J1357+193 were observed every 15\,minutes
for secondary amplitude and phase calibration. The strong calibrator
sources J0423--013, J1058+015, 3C446, and 3C273 were observed at least
once per track for bandpass and secondary flux calibration. Absolute
fluxes were bootstrapped relative to Mars, Uranus, MWC349, or 3C84
(when no planet was available).  Pointing was performed at least every
2--4\,hr on nearby sources, using both radio and optical modes.  The
resulting total calibration is estimated to be accurate within
$\sim$15\%.

All observations were carried out with the previous generation
correlator. The \cco\ line in B1359+154 was centered in the lower
sideband (LSB). All other lines were centered in the upper sideband
(USB), at intermediate frequencies of 2.5\,GHz. Three bands with 15
channels of 31.25\,MHz ($\sim$86--115\,\kms ) width each were centered
on the tuning frequencies. The bands were overlapped by 2 channels to
improve calibration of the correlated dataset, leading to an effective
bandwidth of 1281.25\,MHz ($\sim$3500--4700\,\kms ) per sideband.

For data reduction and analysis, the MIRIAD package was used. The
final plots were created with the GILDAS package. All data were imaged
using `natural' weighting, yielding synthesized beam sizes of
7.3$''$$\times$4.2$''$, 2.5$''$$\times$1.9$''$,
10.1$''$$\times$7.8$''$, and 9.4$''$$\times$6.8$''$ for the \cco\
observations in HE\,0230--2130, HE\,1104--1805, and B1359+154 and the
\dco\ observations of B1359+154, respectively. The final rms noise
values are 0.52, 0.65, 0.97, and 0.71\,mJy\,beam$^{-1}$ over 1114,
450, 344, and 430\,\kms\ (406.25, 156.25, 93.75, and 156.25\,MHz).
Averaging over all line-free data (LSB+USB) in the observations of
B1359+154 (both line setups) yields a beam size of
9.8$''$$\times$7.3$''$ and an rms noise of 0.12\,mJy\,beam$^{-1}$.

\subsection{`Blind' CO Search}

We also used CARMA to observe the very submillimeter-bright,
radio-loud quasar B1938+666, for which the redshift was previously
unknown. Thus, we searched the entire frequency range from 82.4 to
115.3\,GHz for CO line emission, facilitating the large bandwidth of
3708.096\,MHz per sideband (eight bands with 95 channels of 5.208\,MHz
with per sideband, overlapping bands by typically 6\,channels to
reduce sensitivity losses due to bandpass rolloff) of the new
correlator. Observations were carried out for 14\,tracks in the D and
E configurations between 2010 July 05 and September 08, amounting to a
total observing time of 58\,hr (37\,hr on source). A total of 6
(partially overlapping) frequency settings were used to cover the
82.4--115.3\,GHz frequency range. Observations were typically carried
out in pairs of frequency settings with an IF frequency of 3.6\,GHz,
where the LSB of the second setup was used to fill the gap between
sidebands of the first setup (and, vice versa, the USB of the first
setup fills the IF gap of the second setup) to achieve continuous
frequency coverage.

After identification of a CO line, a seventh frequency setup was used
to observe a second CO line, redshifted to $\sim$75.4\,GHz, for
confirmation. This frequency is almost 10\,GHz below the nominal
tuning range of CARMA's 3\,mm receivers (85--116\,GHz). We thus placed
the line in the LSB, using the extended IF range with an IF frequency
of 7.5\,GHz (same bandwidth as above), and placing the local
oscillator (LO) at 82.874\,GHz.\footnote{The USB falls into the
frequency range covered by the previous setups.} With this setup, we
successfully extended the covered frequency range down to
$\sim$74.1\,GHz, with decreasing sensitivity below $\sim$76\,GHz.
These observations were carried out for 3\,tracks in C configuration
between 2010 September 27 and October 10, amounting to a total
observing time of 12\,hr (6.6\,hr on source).

Weather conditions scaled between acceptable and excellent for
observations at 3\,mm wavelengths. The nearby quasar J1849+670 was
observed every 12--20\,minutes for secondary amplitude and phase
calibration. The strong calibrator sources J1751+096, J2015+372,
3C273, 3C345 and 3C454.3 were observed at least once per track for
bandpass and secondary flux calibration. Absolute fluxes were
bootstrapped relative to Neptune and Mars. Pointing was performed at
least every 2--4\,hr on nearby sources, using both radio and optical
modes. The resulting total calibration is estimated to be accurate
within $\sim$15\% (20\% for setup 7).

\begin{deluxetable*}{ l c c c c c c}
\tabletypesize{\scriptsize}
\tablecaption{Observed Line Parameters. \label{t2}}
\tablehead{
target        & line & $z_{\rm CO}$      & $S_{\nu}$ & $\Delta$$v_{\rm FWHM}$ & $I_{\rm CO}$ & $r_{J+1,J}$\tablenotemark{a}\\
              &      &                   & [mJy]        & [\kms ]                & [Jy\,\kms ] }
\startdata    
B1938+666     & \bco & 2.0590$\pm$0.0003 &  9.9$\pm$2.6 & 366$\pm$124            & 3.8$\pm$1.1 & 1.0$\pm$0.3 \\
              & \cco &                   & 16.3$\pm$1.8 & 529$\pm$75             & 9.1$\pm$1.1 & \\
              & \chcn &                  & (3.1$\pm$1.1) & (496$\pm$192)\tablenotemark{b} & (1.62$\pm$0.54)$^\star$ & \\
              & \chco &                  & (1.6$\pm$1.1) & (496$\pm$192)\tablenotemark{b} & (0.86$\pm$0.45)$^\star$ & \\
              & \chnc &                  & (1.9$\pm$1.1) & (496$\pm$192)\tablenotemark{b} & (1.00$\pm$0.46)$^\star$ & \\
              & \ccch &                  & (1.7$\pm$1.1) & (496$\pm$192)\tablenotemark{b} & (0.89$\pm$0.46)$^\star$ & \\
              & \ccn &                   & (1.5$\pm$1.1) &                       & (1.0$\pm$0.7)$^\star$ & \\
HE\,0230--2130 & \cco & 2.1664$\pm$0.0005 & 11.1$\pm$1.5 & 705$\pm$123            & 8.3$\pm$1.2 & \\
HE\,1104--1805 & \cco & 2.3221$\pm$0.0004 & 16.1$\pm$2.4 & 441$\pm$81             & 7.5$\pm$1.2 & \\
B1359+154     & \cco & 3.2399$\pm$0.0003 &  5.6$\pm$1.7 & 198$\pm$92             & 1.2$\pm$0.4 & 1.2$\pm$0.5 \\
              & \dco &                   & 10.0$\pm$1.6 & 237$\pm$47             & 2.5$\pm$0.4 & 
\vspace{-1mm}
\enddata 
\tablecomments{${}$
{$^\star$:\ Considered not detected. Corresponds to 3$\sigma$ upper limits of 1.6 (HCN), 1.4 (HCO$^+$, HNC, C$_2$H), and 2.1\,Jy\,\kms\ (CN).}
}
\tablenotetext{a}{${}$Line brightness temperature ratio CO($J$=n+1$\to$n)/CO($J$=n$\to$n--1), where n=2/3 ($r$=1:\ thermalized).}
\tablenotetext{b}{${}$Fixed to a common linewidth. The width of the C$_2$H line is corrected for broadening due to hyperfine structure.}

\end{deluxetable*}


For data reduction and analysis, the MIRIAD package was used. The
final plots were created with the GILDAS package. All data were imaged
using `natural' weighting, yielding synthesized beam sizes of
2.8$''$$\times$2.3$''$ and 4.5$''$$\times$3.9$''$ at 75.4 and
113.0\,GHz, and 4.5$''$$\times$4.0$''$ over the full, 41.2\,GHz wide
bandpass (with a small gap around 80\,GHz, and excluding line
emission). Due to varying effective exposure time, atmospheric
conditions and receiver noise, the sensitivity varies somewhat over
the full spectral bandpass, in particular toward the edges of the
3\,mm band.  The final rms noise values are 24\,$\mu$Jy\,beam$^{-1}$
over the full bandpass, 0.77\,mJy\,beam$^{-1}$ over 203\,MHz
(808\,\kms ) at 75.4\,GHz, and 0.52\,mJy\,beam$^{-1}$ over 292\,MHz
(773\,\kms ) at 113.0\,GHz.

\section{Results}

\begin{figure*}
\epsscale{1.15}
\plotone{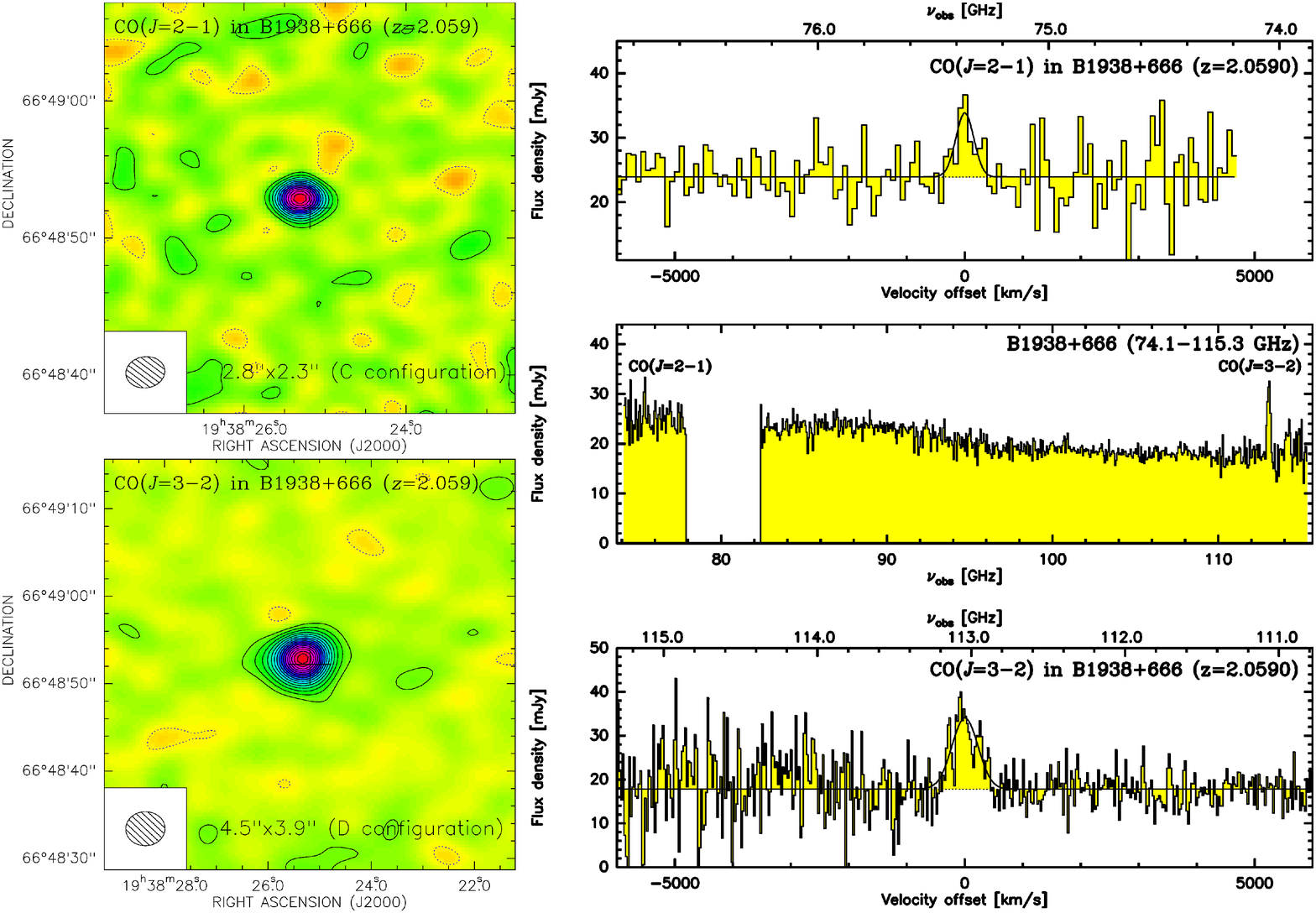}

\caption{CARMA \bco\ and \cco\ maps ({\em left}) and spectra ({\em right}) 
toward B1938+666 ($z$=2.059). Contours are plotted in steps of
3$\sigma$ (1$\sigma$=0.77 and 0.52\,mJy\,beam$^{-1}$ over 808 and
773\,\kms\ (203 and 292\,MHz)), starting at $\pm$3$\sigma$ (no
continuum emission subtracted). The beam sizes are indicated at the
bottom of each map. The crosses indicate the (common) pointing
center. The {\em middle} spectrum (histogram) shows the full covered
frequency range of the redshift search from 74.1 to 115.3\,GHz at a
resolution of 62.496\,MHz. The {\em top} and {\em bottom} spectra
(histograms) of zoomed-in regions around the \bco\ and \cco\ lines are
shown at resolutions of 20.832 and 10.416\,MHz (83 and 28\,\kms\ for
the CO $J$=2$\to$1 and 3$\to$2 lines). The solid black curves are
Gaussian fits to the spectra.
\label{f1b}}
%
\end{figure*}

\subsection{B1938+666}

\subsubsection{Previous Results}

B1938+666 is a radio-selected gravitational lens with a small image
separation of only 0.95$''$ in the radio continuum emission associated
with its radio-loud active galactic nucleus (AGN; Patnaik et al.\
\citeyear{pat92}). The bright radio continuum emission 
(0.577$\pm$0.017\,Jy at 1.4\,GHz) is lensed into at least
3\,components that break up into 7\,subcomponents along a partial
Einstein ring, some of which show polarized emission (King et al.\
\citeyear{kin97}).  The radio components show a relatively steep
spectral index that varies between the different subcomponents
($\alpha_{1.612}^5$=--(1.0--0.5) from 1.612 to 5\,GHz), suggesting a
superposition of flat and steep spectrum emission associated with
lensed core and jet emission.  Its 1.6\,$\mu$m continuum emission is
lensed into a full, almost perfectly symmetric Einstein ring with
0.95$''$ diameter, but only the foreground lensing galaxy is detected
at $i$ band and shorter wavelengths (King et al.\
\citeyear{kin98}). This shows that B1938+666 has properties that are
clearly different from optically-selected, gravitationally lensed
quasars.  The lensing galaxy has a spectroscopic redshift of
$z$=0.881, suggesting $z$$>$1.7 for B1938+666 based on the lensing
configuration, with a most likely redshift of $z$=2.8 (Tonry \&
Kochanek \citeyear{tk00}). Until now, the redshift of B1938+666 was
highly uncertain. In addition to its radio-loud AGN, B1938+666 also
hosts a luminous dust bump, with $S_{450\,{\mu}{\rm
m}}$=126$\pm$22\,mJy, $S_{850\,\mu{\rm m}}$=34.6$\pm$2.0\,mJy, and
$S_{1.3\,{\rm mm}}$=14.7$\pm$2.0\,mJy (Barvainis \& Ivison
\citeyear{bi02}). Correcting for synchrotron emission from the AGN,
Barvainis \& Ivison (\citeyear{bi02}) estimate that 32.0$\pm$6.7\,mJy
of the 850\,$\mu$m emission are due to dust heating, likely dominated
by star formation. They also find a model-based lensing magnification
factor of $\mu_{\rm L}$=173.  Its extreme submillimeter brightness,
dust obscuration in the optical and small Einstein radius (making
optical spectroscopic redshift searches unfeasible) make B1938+666 an
ideal candidate to attempt a `blind' CO redshift search.

\subsubsection{New Results:\ `Blind' CO Redshift and Line Emission}

Scanning the whole 3\,mm band from 82.4--115.3\,GHz, we have
identified a single CO line in B1938+666 at 113.0425\,GHz. The lack of
detecting a second, lower-$J$ CO line at 82.4$<$$z$$<$113.0\,GHz rules
out any candidate redshifts at $z$$>$3.0. Assuming a lower limit of
$z$=1.7 as suggested by the lensing configuration, this leaves \cco\
emission at $z$=2.059 as the only likely explanation for the detected
line emission. We confirmed this conclusion through a subsequent
detection of \bco\ emission at 75.3638\,GHz.

In Fig.~\ref{f1b}, the full covered spectral range is shown, along
with higher spectral resolution, zoomed-in spectra of the emission
lines, and emission line maps (continuum not subtracted). As both
redshifted CO lines are close to the edges of the 3\,mm band, the
noise across the zoomed-in spectra is not flat, and substantially
increases below $\sim$75\,GHz and above $\sim$114\,GHz. From Gaussian
fitting to the line profiles, we derive line peak flux densities of
$S_\nu$=9.9$\pm$2.6 and 16.3$\pm$2.8\,mJy at FWHM widths of
$\Delta$$v_{\rm FWHM}$=366$\pm$124 and 529$\pm$75\,\kms\ for the
\bco\ and \cco\ lines, respectively (see Tab.~\ref{t2}). The line
widths are consistent within the errors. A comparison of the line
profiles may suggest more complex structure than a single Gaussian,
but more sensitive observations are required to investigate this in
more detail. The observed-frame peak velocities correspond to a median
redshift of $z_{\rm CO}$=2.0590$\pm$0.0003. We find
velocity-integrated \bco\ and \cco\ line fluxes of $I_{\rm
CO}$=3.8$\pm$1.1 and 9.1$\pm$1.1\,Jy\,\kms. The line brightness
temperature ratio of $r_{32}$=1.0$\pm$0.3 is consistent with optically
thick, thermalized emission within the errors, consistent with what is
found in other high-$z$ quasars (e.g., Riechers et al.\
\citeyear{rie06}).

We derive a lensing-corrected CO line luminosity of $L'_{\rm
CO}$=1.3$\times$10$^9$\,($\mu_{\rm
L}$/173)$^{-1}$\,K\,\kms\,pc$^2$. Assuming a conversion factor of
$\alpha_{\rm CO}$=0.8\,\msol\,(K\,\kms\,pc$^2$)$^{-1}$ from $L'_{\rm
CO}$ to gas mass for ultra-luminous infrared galaxies (ULIRGs; Downes
\& Solomon \citeyear{ds98}), we find \\
$M_{\rm gas}$=1.0$\times$10$^9$\,($\mu_{\rm
L}$/173)$^{-1}$\,($\alpha_{\rm CO}$/0.8)\,\msol\ (see
Tab.~\ref{t3}). In their analysis, Barvainis \& Ivison
(\citeyear{bi02}) introduce a cutoff in $\mu_{\rm L}$ of a factor of
20, which may be more appropriate if the gas reservoir extends over a
$\gg$100\,pc size region. Under this assumption, $L'_{\rm CO}$ and
$M_{\rm gas}$ would be $\sim$8$\times$ higher, but still at the low
end of observed values at high $z$.

\begin{figure}
\epsscale{1.15}
\plotone{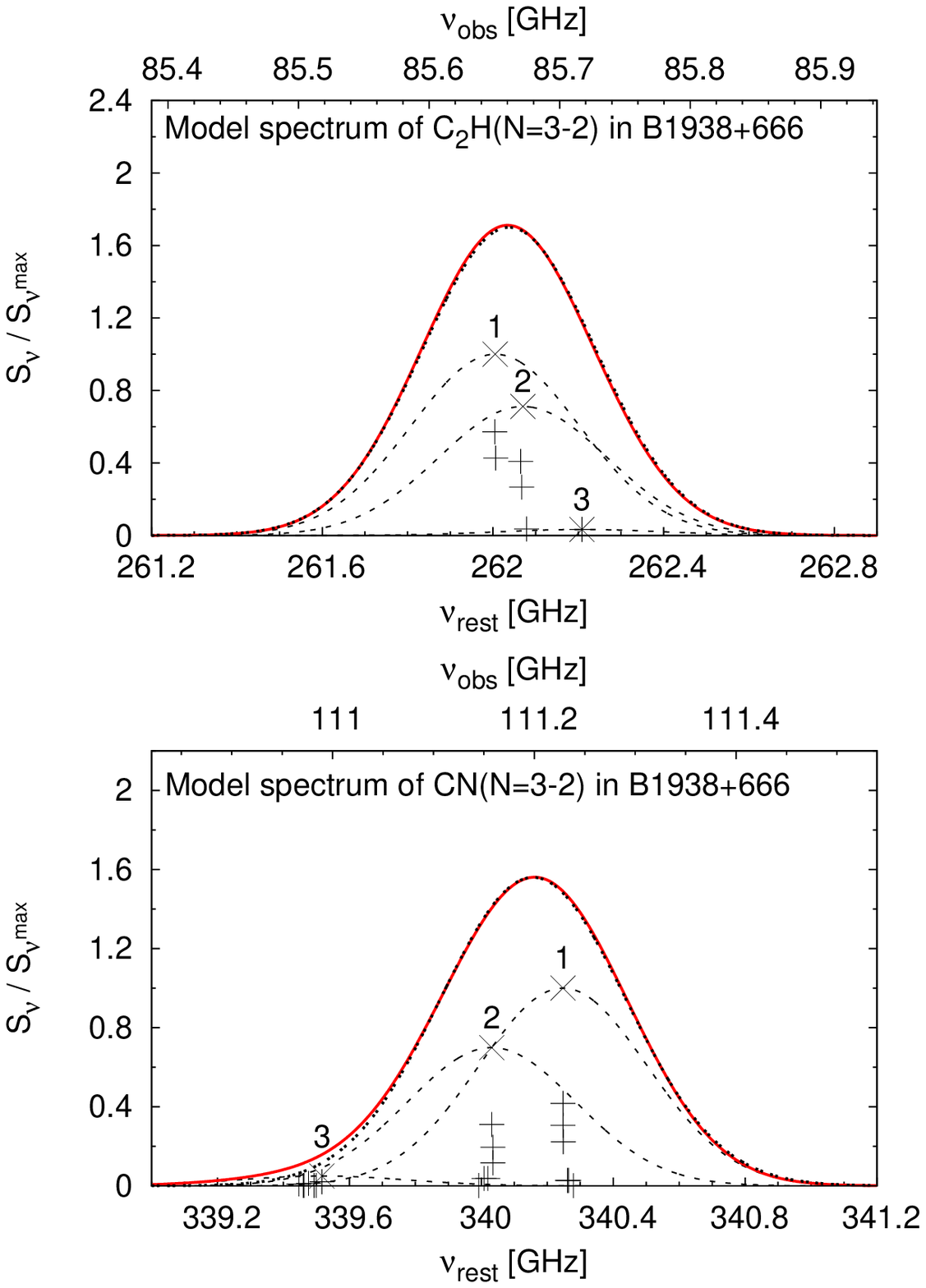}
\vspace{-4mm}

\caption{Model spectra of the C$_2$H($N$=3$\to$2) and \ccn\ emission 
toward B1938+666. The relative intensities are computed in local
thermodynamic equilibrium (see Riechers et al.\ \citeyear{rie07a},
\citeyear{rie09a}, for details on the computation). The plus signs
indicate all hyperfine components, for each of which the same line
width as for the \cco\ line is assumed. The crosses and dashed lines
indicate the peak intensities and profiles of the three main
subcomponents (labeled ``1'' to ``3''), summed over hyperfine
components close in frequency.  The vertical axes indicate the
predicted intensity, normalized to the brightest subcomponent.  The
two brightest of these subcomponents for each line are indicated in
Fig.~\ref{f11}. The solid lines are the profiles summed over all
components. The thick dotted line indicates single Gaussian profiles
fitted to the summed profiles, and have widths that are by $\sim$2.6\%
(C$_2$H) and $\sim$10\% (CN) larger than that of the individual
components (and are used to fit the observations to minimize the
number of fitting parameters).
\label{f12}}
%
\end{figure}

\subsubsection{New Results:\ Constraints on HCN, HCO$^+$, HNC, C$_2$H, and CN Line Emission}

\begin{figure}
\epsscale{1.15}
\plotone{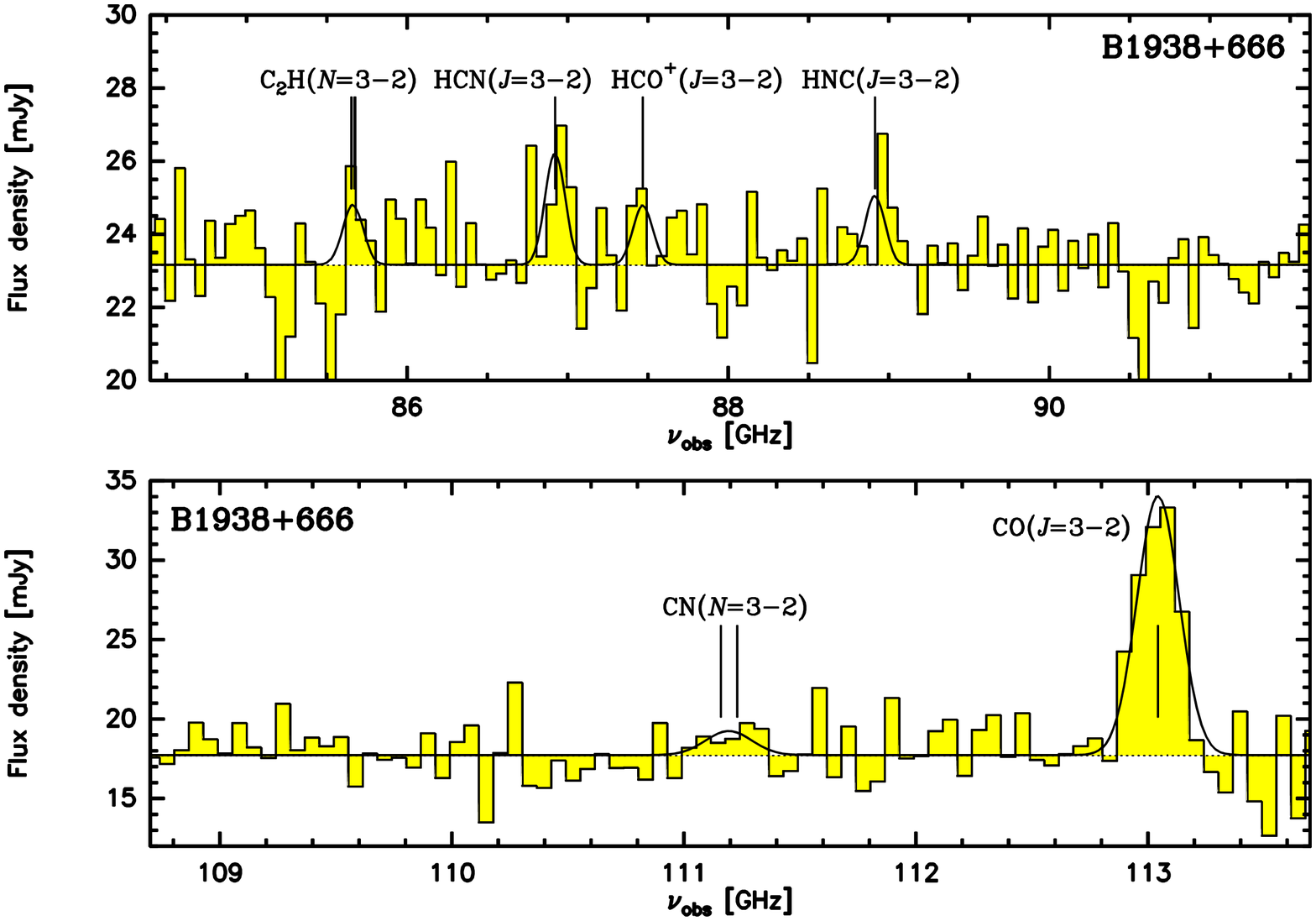}

\caption{A Search for \chcn, \chco, \chnc, C$_2$H($N$=3$\to$2), 
and \ccn\ emission toward B1938+666 ($z$=2.059). The spectra
(histogram) are shown at the same resolution as in the {\em middle}
panel of Fig.~\ref{f1b}. {\em Top}: The solid black curve shows a
simultaneous Gaussian fit to all four lines with a common line width
(corrected for C$_2$H hyperfine structure; see Fig.~\ref{f12}) and the
underlying continuum emission, with the redshift fixed to
$z$=2.0590. The peak frequencies of the \chcn, \chco, and \chnc\ lines
as well as the main hyperfine structure components of the
C$_2$H($N$=3$\to$2) line are indicated by the vertical lines.  {\em
Bottom}: The solid black curve shows a simultaneous Gaussian fit to
both lines with a common line width (corrected for CN hyperfine
structure; see Fig.~\ref{f12}) and the underlying continuum
emission. The peak frequencies of the main hyperfine components of
\ccn\ and that of the
\cco\ line are indicated by the vertical lines.
\label{f11}}
%
\end{figure}

Based on the firm CO redshift, we went back to the spectrum to search
for emission from dense molecular gas tracers. Fixing the redshift to
$z$=2.0590 and all four lines to a common width (the width of the
C$_2$H line was corrected up by 2.6\% to account for hyperfine
structure; see Fig.~\ref{f12}; and Riechers et al.\ \citeyear{rie07a},
\citeyear{rie09a}, for more details on the fitting procedure), we
simultaneously fitted Gaussian line profiles at the redshifted
frequencies of \chcn, \chco, \chnc, and \ccch\ emission,\footnote{The
rest frequencies of the HCN, HCO$^+$, and \chnc\ lines are $\nu_{\rm
rest}$=265.886180, 267.557619, and 271.981142\,GHz. The \ccch\ line
has six spectrally distinguishable hyperfine components at $\nu_{\rm
rest}$=262.004227--262.208439\,GHz (see Fig.~\ref{f12}).} adding one
free parameter to fit the underlying continuum (see
Fig.~\ref{f11}). We find $S_\nu$=3.1$\pm$1.1, 1.6$\pm$1.1,
1.9$\pm$1.1, and 1.7$\pm$1.1\,mJy at a common $\Delta$$v_{\rm
FWHM}$=496$\pm$192\,\kms\ for the HCN, HCO$^+$, and \chnc\ and \ccch\
lines, respectively. This corresponds to $I_{\rm HCN}$=1.62$\pm$0.54,
$I_{\rm HCO^+}$=0.86$\pm$0.45, $I_{\rm HNC}$=1.00$\pm$0.46, and
$I_{\rm C_2H}$=0.89$\pm$0.46\,Jy\,\kms, respectively. We also searched
for CN emission, simultaneously fitting the \ccn\ and \cco\ lines and
the underlying continuum emission (see Fig.~\ref{f11}). The width of
the CN line was corrected up by 10\% to account for hyperfine
structure (see Fig.~\ref{f12}).\footnote{The
\ccn\ line has 19 spectrally distinguishable hyperfine components at
$\nu_{\rm rest}$=339.446777--340.279166\,GHz (see Fig.~\ref{f12}).} The
fit yields $S_\nu$=1.5$\pm$1.1\,mJy. corresponding to $I_{\rm
HCN}$=1.0$\pm$0.7\,Jy\,\kms.

The \chcn\ line is marginally detected at 3.0$\sigma$ level, but we
conservatively treat it as an upper limit in the following. The other
four lines yield 1.4--2.2$\sigma$ signals, and thus are not
detected. Translating their fluxes to 3$\sigma$ upper limtis yields
$I$$<$1.4 (HCO$^+$, HNC, C$_2$H) and $<$2.1\,Jy\,\kms\ (CN). Given
that the simultaneously fitted HCN, HCO$^+$, HNC, and C$_2$H lines all
yield positive signals at 1.9--3.0$\sigma$ level, we attempted to
stack their spectra. This yields a weighted average flux of
1.05$\pm$0.24\,Jy\,\kms, corresponding to a signal of 4.4$\sigma$
significance. This stacked signal corresponds to 19\%$\pm$5\% of the
\cco\ luminosity (i.e., not corrected for line excitation). This is 
consistent with the median HCN/CO $L'$ ratios (a measure of the dense
gas fraction in galaxies) in nearby ULIRGs and $z$$>$2 quasars (e.g.,
Gao et al.\ \citeyear{gao07}; Riechers et al.\ \citeyear{rie07b}).

\subsubsection{New Results:\ Continuum Emission}

\begin{figure*}
\epsscale{1.15}
\plotone{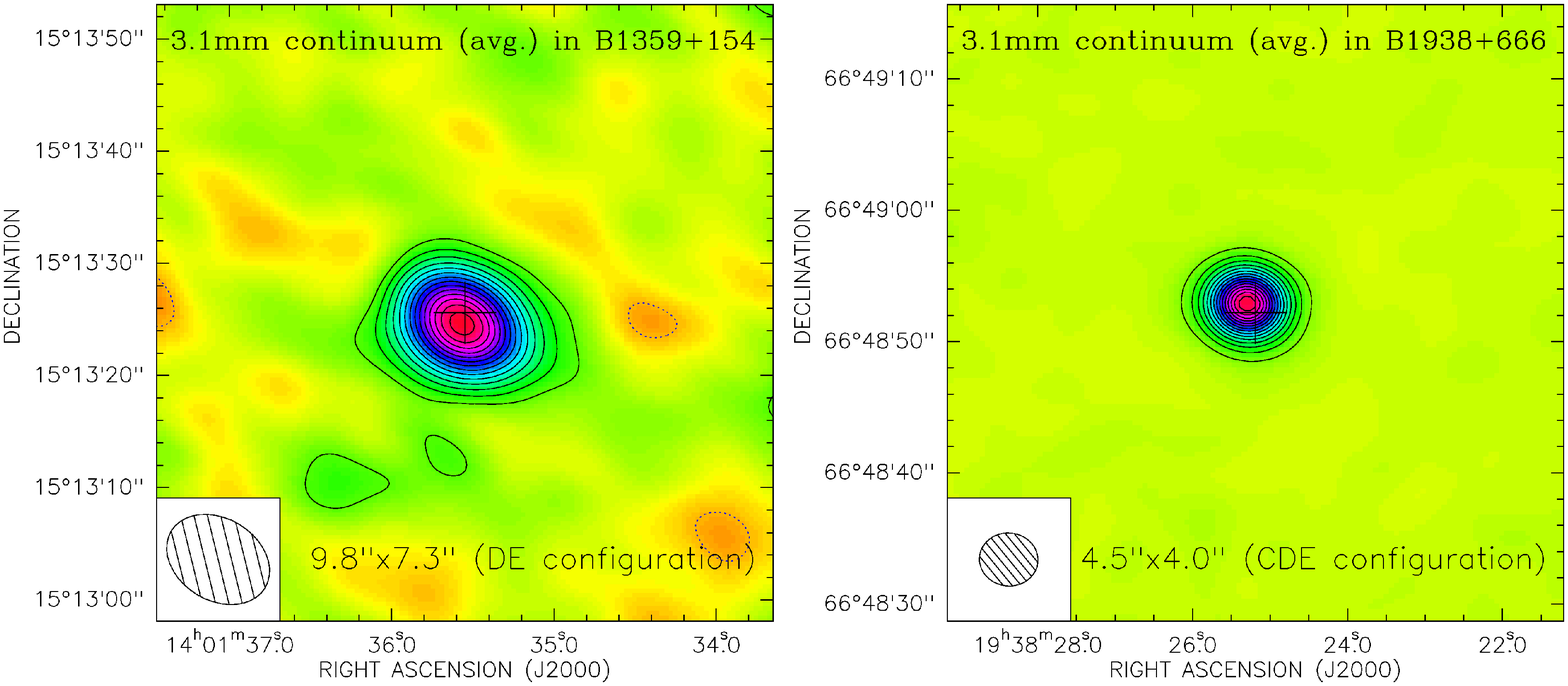}

\caption{3.1\,mm continuum emission in B1359+154 and B1938+666, averaged 
over all line-free data in the lower and upper sidebands.  The beam
sizes are indicated at the bottom. The crosses indicate the same
position as in Figs.~\ref{f1b} and \ref{f1}.  {\em Left}:\ Contours
are plotted in steps of 2$\sigma$ (1$\sigma$=0.12\,mJy\,beam$^{-1}$),
starting at $\pm$3$\sigma$. {\em Right}:\ Contours are plotted in
steps of 50$\sigma$ (1$\sigma$=24\,$\mu$Jy\,beam$^{-1}$), starting at
$\pm$50$\sigma$.\label{f2}}

\end{figure*}

As shown in Fig.~\ref{f1b}, we detect continuum emission toward
B1938+666 over the entire observed wavelength range. Simultaneous fits
to the line and continuum emission suggest continuum strengths of
23.89$\pm$0.32 and 17.76$\pm$0.26\,mJy below the \bco\ and \cco\
lines, and 23.15$\pm$0.15\,mJy below the \chcn, \chco, \chnc, and
\ccch\ lines. The strength and slope of the emission is consistent
with synchrotron emission from the radio-loud AGN. The spectrum may
suggest a more complex continuum slope than a simple power
law. However, due to variability of the phase calibrator over the
4\,month course of the observations, there is a residual uncertainty
in the flux calibration between the different frequency settings. We
thus conclude that the continuum slope is consistent with a power law
within the errors. By averaging the line-free channels over the entire
observed frequency range and fitting a 2-dimensional Gaussian to the
$u-v$ data, we find a 3.1\,mm continuum flux of 20.84$\pm$0.14\,mJy.
In Fig.~\ref{f2}, a map of the averaged continuum emission is shown.
From the fit, we also find a continuum size of
(1.48$''$$\pm$0.05$''$)$\times$(1.10$''$$\pm$0.05$''$) at a position
angle of 85$^\circ$$\pm$5$^\circ$. Accounting for interferometric
seeing (which is not part of the above errors), this is consistent
with the 0.95$''$ diameter of the Einstein ring, and the continuum
emission to be dominated by the radio-loud AGN.

\subsubsection{Origin of the CO/Continuum Emission}

\begin{figure*}
\epsscale{1.15}
\plotone{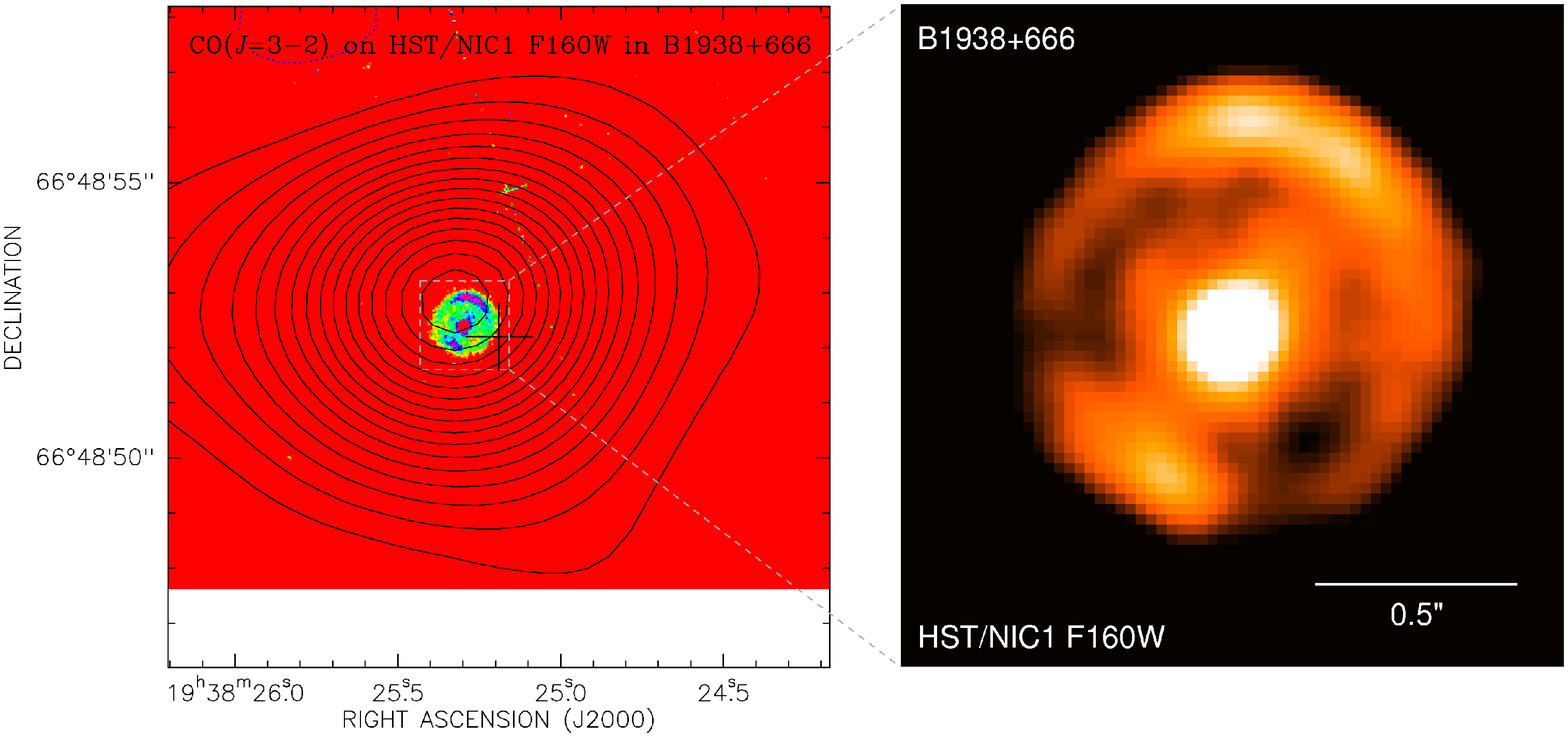}

\caption{Overlay of \cco\ (contours) and  1.6\,$\mu$m (rest-frame 523\,nm) 
continuum emission ({\em left}; {\em HST}/NICMOS1; image not cleaned;
the background source is not detected at 555 and 814\,nm), and
zoomed-in intensity map of the 1.6\,$\mu$m emission ({\em right};
smoothed with a 0.1$''$ Gaussian kernel) in B1938+666. The cross
indicates the same position as in Fig.~\ref{f1b}.  The source is
lensed into a full Einstein ring, and the central spot is the
foreground lensing galaxy.
\label{f6}}

\end{figure*}

\begin{deluxetable*}{l c c c c c c c c c}
\tabletypesize{\scriptsize}
\tablecaption{Luminosities, gas masses and star-formation rates (SFRs).\label{t3}}
\tablehead{
Source          & $D_{\rm L}$ & $\mu_{\rm L}$\tablenotemark{a} & $L_{\rm FIR}$\tablenotemark{b}$^,$\tablenotemark{c} & $L'_{\rm CO}$\tablenotemark{b}$^,$\tablenotemark{d} & $L_{\rm FIR}$/$L'_{\rm CO}$ & $M_{\rm gas}({\rm H_2})$\tablenotemark{e} & SFR\tablenotemark{f} & $\tau_{\rm dep}$\tablenotemark{g} \\
                & [Gpc]       &     & [10$^{12}\,$L$_{\odot}$]       & [10$^{10}\,$K \kms pc$^2$] & [L$_{\odot}$/K \kms pc$^2$]        & [10$^{10}$\,M$_{\odot}$] & [M$_{\odot}$\,yr$^{-1}$] & [Myr] }
\startdata
B1938+666     & 16.31       & 173  & 33/0.19       & (22.0$\pm$2.5)/0.13       &  150                             & 0.10                    &  30 & 35 \\
HE\,0230--2130 & 17.36       & 14.5 & 21/1.5        & (21.9$\pm$3.1)/1.5        &  100                             & 1.2                     & 220 & 55 \\
HE\,1104--1805 & 18.91       & 10.8 & 16/1.5        & (22.4$\pm$3.4)/2.1        &   70                             & 1.7                     & 220 & 75 \\
B1359+154     & 28.36       & 118  & 11/0.093      &  (7.7$\pm$1.2)/0.065      &  140                             & 0.052                   &  14 & 40 
\vspace{-1mm}
\enddata
\tablenotetext{a}{Adopted from Barvainis \& Ivison \citeyear{bi02}.}
\tablenotetext{b}{Apparent luminosities (not corrected for lensing)/intrinsic luminosities (lensing--corrected).}
\tablenotetext{c}{$L_{\rm FIR}$ calculated assuming similar SED shapes as for the Cloverleaf quasar (Wei\ss\ et al.\ \citeyear{wei03}).}
\tablenotetext{d}{Extrapolated to \aco\ luminosity by correcting for excitation, based on excitation modeling of high-$z$ quasars (Riechers et al.\ \citeyear{rie06}, \citeyear{rie09a}). Corrections applied:\ 0\% for \bco, 2\% for \cco, 6\% for \dco.}
\tablenotetext{e}{Assuming a conversion factor of $\alpha = 0.8$\,M$_{\odot}\,$/K \kms pc$^2$ from $L'_{\rm CO(1-0)}$ to $M_{\rm gas}({\rm H_2})$ as appropriate for ULIRGs (see Downes \& Solomon \citeyear{ds98}).}
\tablenotetext{f}{Assuming (Kennicutt \citeyear{ken98a}; \citeyear{ken98b}):\ SFR[M$_{\odot}$\,yr$^{-1}$]$ = 1.5 \times 10^{-10}\,L_{\rm FIR}$[L$_{\odot}$], 
i.e., $\delta_{\rm MF} \delta_{\rm SB} = 1.5$ following the notation of Omont et al.\ \citeyear{omo01}:\ $\delta_{\rm MF}$ describes the 
dependence on the mass function of the stellar population, $\delta_{\rm SB}$ gives the fraction of $L_{\rm FIR}$ that is actually 
powered by the starburst and not the AGN.}
\tablenotetext{g}{Gas depletion timescales, defined as $\tau_{\rm dep}$=$M_{\rm gas}$/SFR.}

\end{deluxetable*}


In Figure \ref{f6}, an overlay of the \cco\ and underlying continuum
emission on top of observed-frame 1.6\,$\mu$m continuum emission in
B1938+666 is shown ({\em Hubble Space Telescope (HST)} image from King
et al.\ \citeyear{kin98}; all {\em HST} images are data products from
the Hubble Legacy Archive).\footnote{\tt http://hla.stsci.edu/} The
peak of the CO/continuum emission is consistent with the brightest
rest-frame optical (523\,nm) emission region along the Einstein ring.
This suggests that the optical and (rest frame) (sub-)millimeter
emission (observed frame 2.6--4.0\,mm) are likely cospatial, with the
optical emission being dominated by stellar light above the Balmer
break. The lack of pointlike images along the Einstein ring suggest
that the radio-loud AGN is optically obscured (or, alternatively, has
low optical luminosity). The strength and spectral slope of the
rest-frame submillimeter continuum emission make it unlikely that the
AGN is much less magnified than its host galaxy. Thus, B1938+666
likely hosts a type-2 AGN along with its starburst.

\begin{figure}
\epsscale{1.15}
\plotone{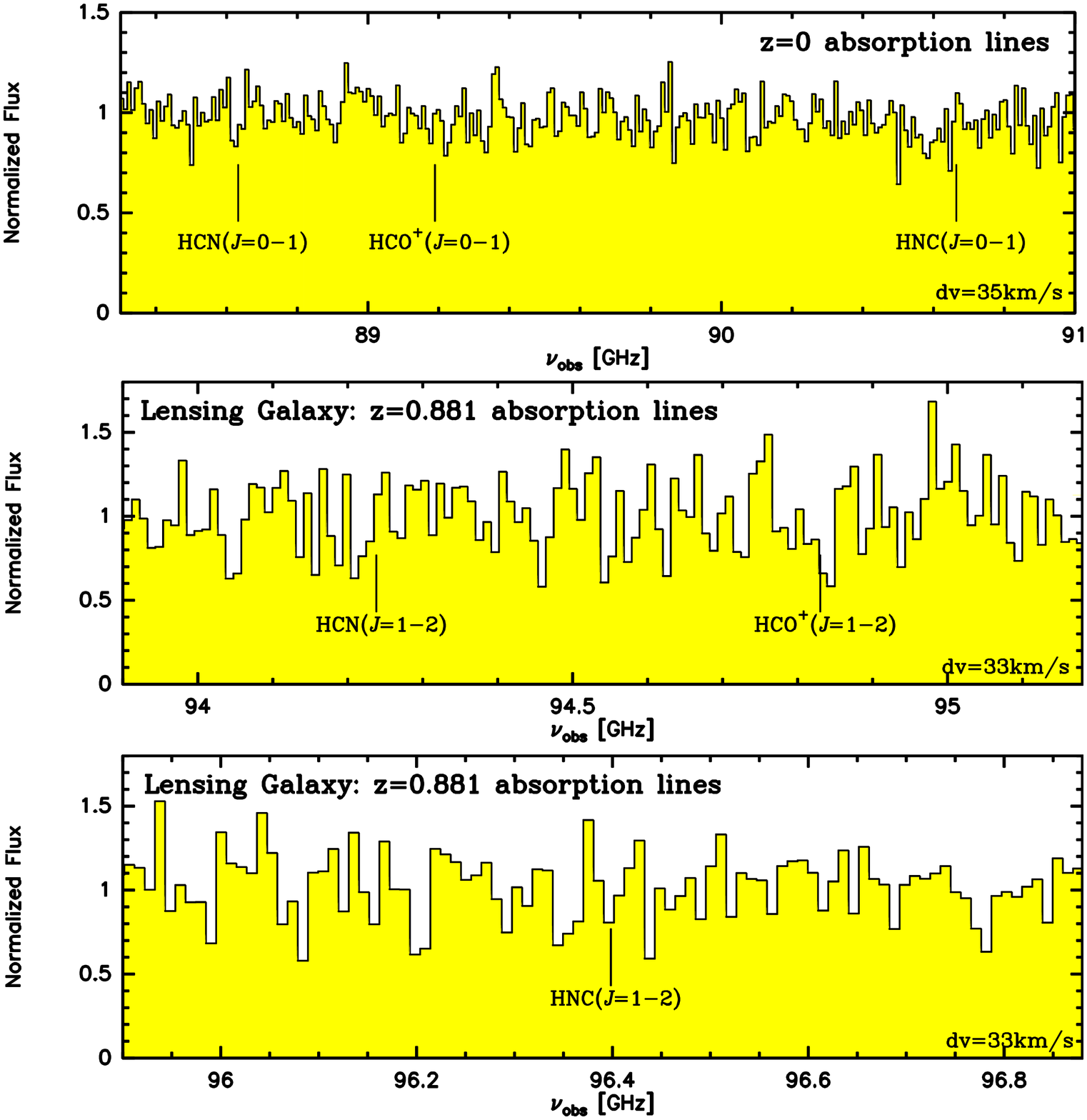}

\caption{Absorption line searches toward the continuum of B1938+666. 
The {\em top panel} shows a frequency region covering the \ahcni,
\ahcoi, and \ahnci\ lines at $z$=0 (i.e., Galactic regions) at a 
velocity resolution of 35\,\kms\ (10.416\,MHz). The {\em middle} and
{\em bottom} panels show frequency regions covering the \bhcni,
\bhcoi, and \bhnci\ lines at $z$=0.881, i.e., the redshift
of the lensing galaxy, at a velocity resolution of 33\,\kms\
(10.416\,MHz). All spectra are normalized to an arbitrary flux scale.
\label{f10}}
%
\end{figure}

\subsubsection{Absorption Line Searches}

Given the high 3\,mm continuum flux of B1938+666, we have searched
the spectrum for absorption lines from foreground sources. In
Fig.~\ref{f10}, three spectral regions are shown that cover the
absorption line frequencies of \ahcni, \ahcoi, and \ahnci\ in the
Milky Way (i.e., at $z$=0), and of \bhcni, \bhcoi, and \bhnci\ at
$z$=0.881, the redshift of the lensing galaxy. HCN, HCO$^+$, or HNC
are abundant molecules typically found in dense regions, and typically
produce some of the deepest absorption features (e.g., Wiklind \&
Combes \citeyear{wk95}, \citeyear{wk96}). The spectra are normalized
to an arbitrary flux scale, and re-binned to 10.416\,MHz (35 and
33\,\kms ), to enable searches for narrow lines from compact molecular
structures. At this resolution, the continuum emission is detected in
every single velocity channel, but at relatively moderate
signal-to-noise ratio. We do not detect any absorption lines. Assuming
a velocity width that is at least comparable to the resolution of
these spectra, this suggests that $<$20\% of the continuum emission of
B1938+666 are absorbed by Galactic or nearby extragalactic foreground
structures, and $<$40\% are absorbed by molecular gas (HCN, HCO$^+$,
or HNC) in the lensing galaxy.

\subsection{HE\,0230--2130:\ The Southern Cloverleaf}

\subsubsection{Previous Results}

HE\,0230--2130 is an optically selected, quadruply lensed quasar at an
optical redshift of $z_{\rm opt}$=2.163$\pm$0.003 with a maximum image
separation of 2.15$''$ (Wisotzki et al.\ \citeyear{wis99}; Anguita et
al.\ \citeyear{ang08}), similar to the well-known Cloverleaf quasar
(Magain et al.\ \citeyear{mag88}). It is lensed by two galaxies at
almost identical redshift ($z_{\rm G1}$=0.523$\pm$0.001, $z_{\rm
G2}$=0.526$\pm$0.002; Eigenbrod et al.\ \citeyear{eig06}).  This
`Southern Cloverleaf' is radio-quiet, but has a luminous dust bump,
with $S_{450\,{\mu}{\rm m}}$=77$\pm$13\,mJy and $S_{850\,\mu{\rm
m}}$=21.0$\pm$2.2\,mJy. These fluxes are $\sim$35\% of those found for
the Cloverleaf, and consistent with the same spectral slope (Barvainis
\& Ivison \citeyear{bi02}). Barvainis \& Ivison find a model-based
lensing magnification factor of $\mu_{\rm L}$=14.5, which is similar
to that of the Cloverleaf ($\mu_{\rm L}$=11; Venturini \& Solomon
\citeyear{vs03}). HE\,0230--2130 also exhibits luminous emission from
polycyclic aromatic hydrocarbons (PAHs; Lutz et al.\
\citeyear{lut08}). Together, these properties suggest the presence of
a substantial molecular gas reservoir.

\begin{figure*}
\epsscale{.77}
\plotone{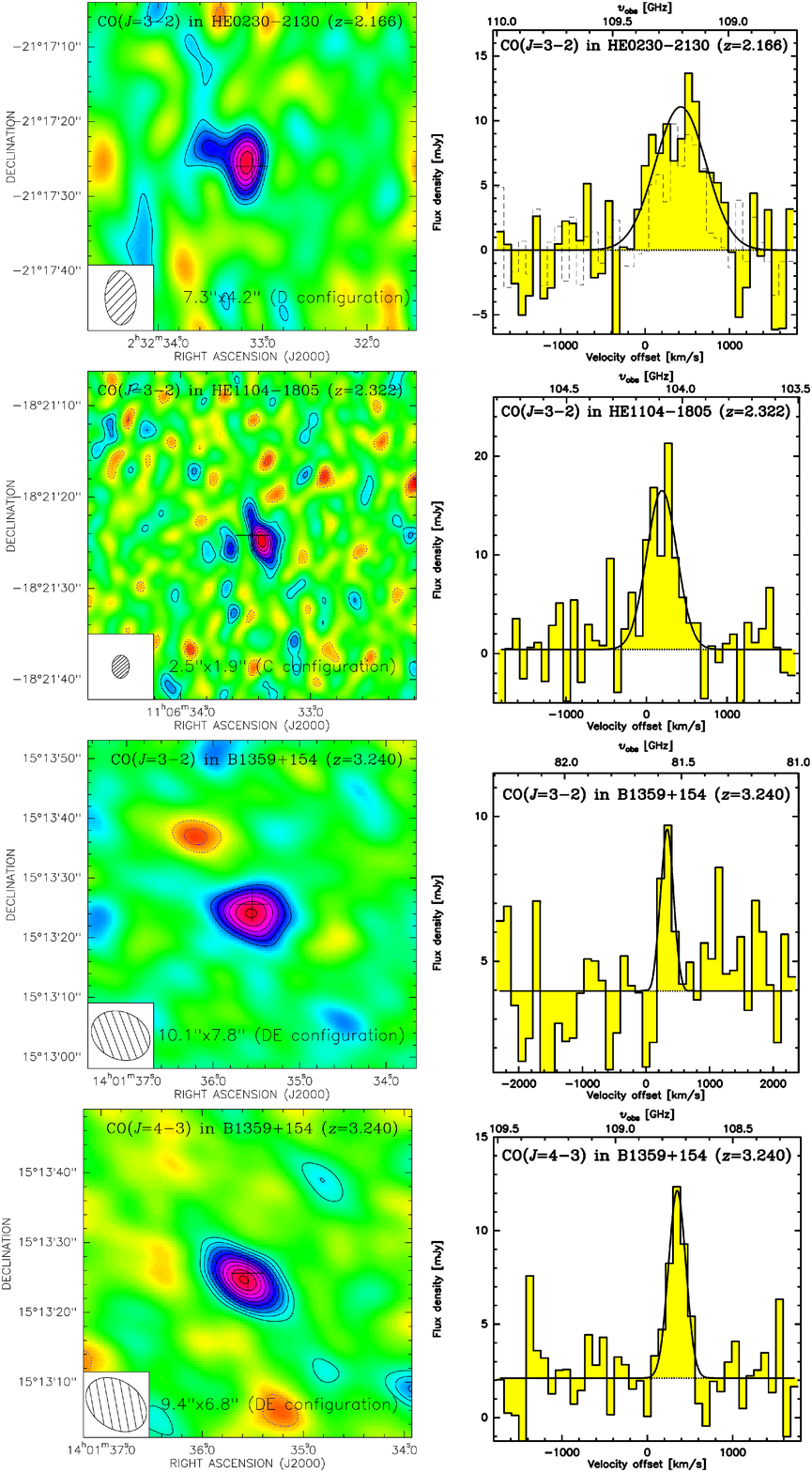}

\caption{CARMA \cco\ and \dco\ maps ({\em left}) and spectra ({\em right}) 
toward HE\,0230--2130 ($z$=2.166), HE\,1104--1805 ($z$=2.322), and
B1359+154 ($z$=3.240). Contours are plotted in steps of 1$\sigma$
(0.52, 0.65, 0.97, and 0.71\,mJy\,beam$^{-1}$ over 1114, 450, 344, and
430\,\kms\ (406.25, 156.25, 93.75, and 156.25\,MHz)), starting at
$\pm$3$\sigma$, $\pm$2$\sigma$, $\pm$3$\sigma$, and $\pm$3$\sigma$ (no
continuum emission subtracted). The beam sizes are indicated at the
bottom of each map. The crosses indicate the pointing centers. The
spectra (histograms) are shown at a resolution of 31.25\,MHz (86, 90,
115, and 86\,\kms ). The solid black curves are Gaussian fits to the
spectra. The dashed histogram in the {\em top right} panel indicates
the peak position spectrum only.
\label{f1}}
%
\end{figure*}

\subsubsection{New Results}

We detect luminous, marginally resolved \cco\ emission toward
HE\,0230--2130 at $>$9$\sigma$ significance (Fig.~\ref{f1}). From
Gaussian fitting to the line profile, we find
$S_\nu$=11.1$\pm$1.5\,mJy and $\Delta$$v_{\rm
FWHM}$=705$\pm$123\,\kms, corresponding to $I_{\rm
CO}$=8.3$\pm$1.2\,Jy\,\kms (see Tab.~\ref{t2}). We do not detect the
underlying continuum down to a 2$\sigma$ limit of $<$1.5\,mJy. The CO
line emission yields a host galaxy redshift of $z_{\rm
CO}$=2.1664$\pm$0.0005, which is consistent with the optical redshift
of the quasar within the errors (d$z$=$\|$$z_{\rm CO}$--$z_{\rm
opt}$$\|$$\simeq$0.003).

\begin{figure}
\epsscale{1.15}
\plotone{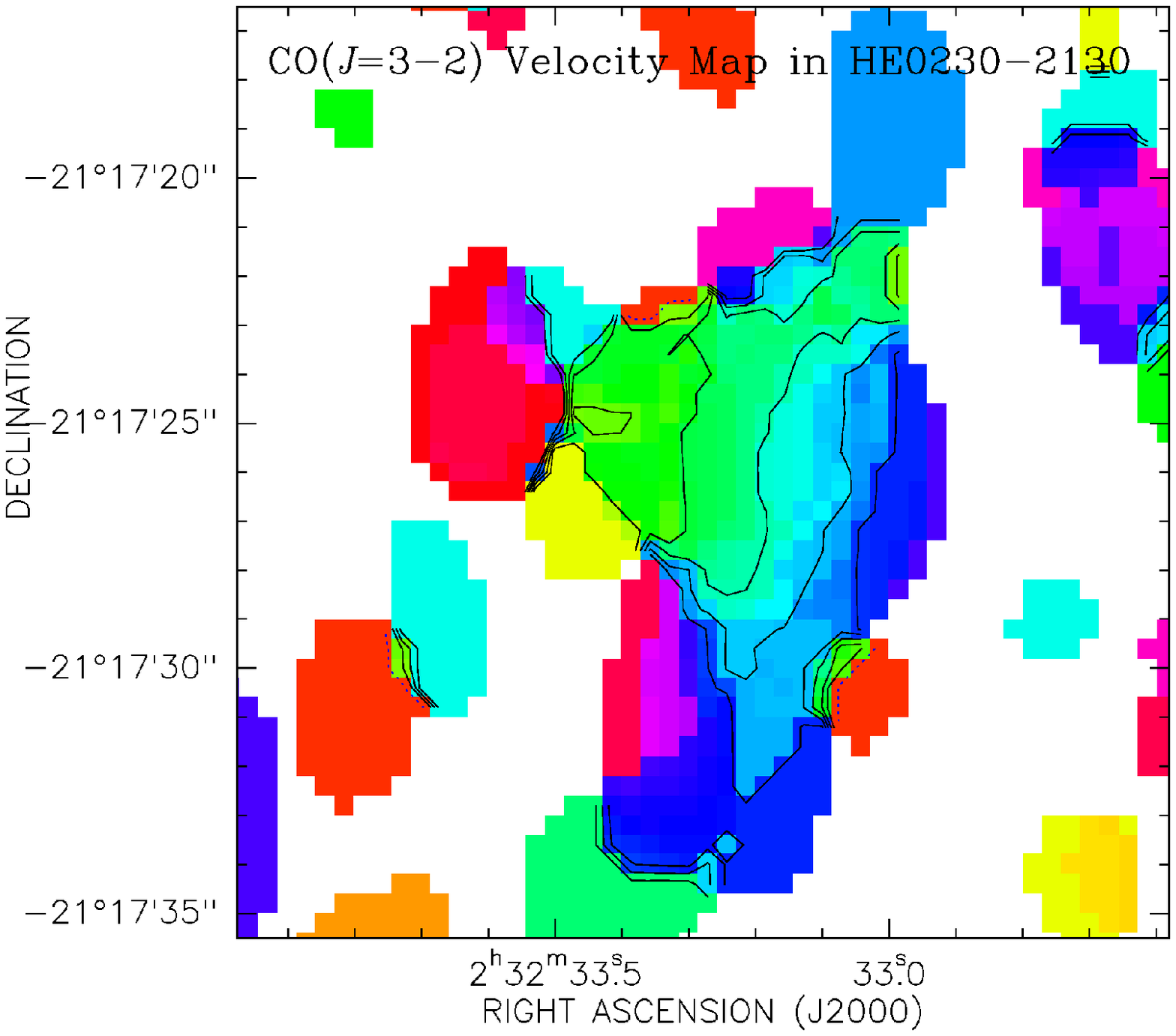}

\caption{First moment map of the \cco\ emission in HE\,0230--2130 shown 
in Fig.~\ref{f1}. Velocity contours are in steps of $\Delta$$v$=100\,\kms.
\label{f3}}
%
\end{figure}

The CO line emission in HE\,0230--2130 appears resolved both spatially
and in velocity. There is a clear difference between the CO line width
at peak position (dashed spectrum in Fig.~\ref{f1}) and the spatially
integrated line width.  In Figure \ref{f3}, a first moment map is shown
(clipped below 1$\sigma$ in the individual velocity channels). Despite
the moderate signal-to-noise ratio of this map, a clear velocity
gradient is apparent. Accounting for beam convolution, the spatial and
velocity structure is consistent with the lensed size of the source as
defined by the maximum images separation of 2.15$''$ within the
uncertainties.

Assuming a 2\% correction for subthermal excitation of the \cco\
line\footnote{This model-based, small correction factor is not
statistically significant, but applied for consistency with estimates
based on higher-$J$ lines, as described below. Within the
uncertainties of our measurements, the \cco\ lines in our targets are
consistent with thermal excitation.}  (based on CO line radiative
transfer modeling of high-$z$ quasars, e.g., Riechers et al.\
\citeyear{rie06}; \citeyear{rie09b}) we derive $L'_{\rm
CO}$=1.5$\times$10$^{10}$\,($\mu_{\rm
L}$/14.5)$^{-1}$\,K\,\kms\,pc$^2$, and $M_{\rm
gas}$=1.2$\times$10$^{10}$\,($\mu_{\rm L}$/14.5)$^{-1}$\,($\alpha_{\rm
CO}$/0.8)\,\msol\ (see Tab.~\ref{t3}).

\subsection{HE\,1104--1805}

\begin{figure*}
\epsscale{1.15}
\plotone{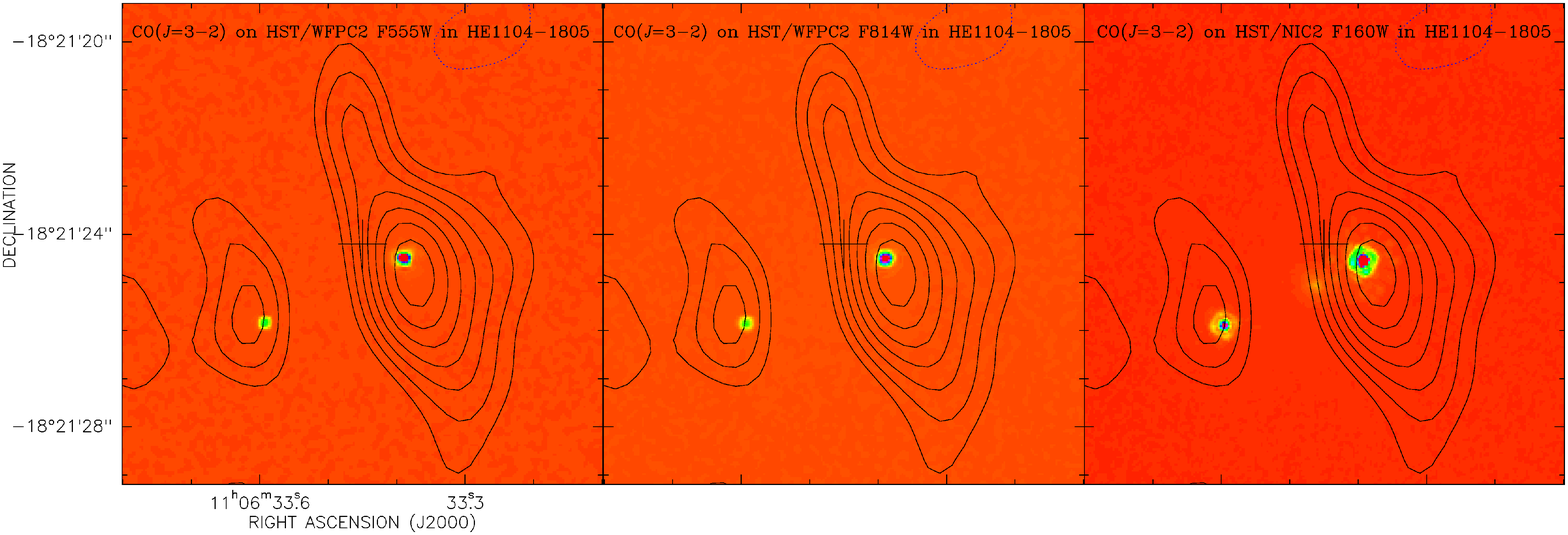}

\caption{Overlays of \cco\ (contours) and 555\,nm, 814\,nm, and 1.6\,$\mu$m 
(rest-frame 167, 245, and 482\,nm) continuum emission ({\em HST}/WFPC2
and NICMOS2, {\em left} to {\em right}; images not cleaned) in
HE\,1104--1805. The cross indicates the same position as in
Fig.~\ref{f1}. The two brightest spots are the lens images of the
background quasar, and the fainter spot in the middle is the
foreground lensing galaxy.\label{f4}}
%
\end{figure*}

\subsubsection{Previous Results}

HE\,1104--1805 is an optically selected double image quasar with a
lens image separation of 3.19$''$ at an optical redshift of $z_{\rm
opt}$=2.3192$\pm$0.0007 (Wisotzki et al.\ \citeyear{wis95}; Sulentic
et al.\ \citeyear{sul06}). It is lensed by a galaxy at $z_{\rm
G}$=0.729$\pm$0.001 (Lidman et al.\ \citeyear{lid00}).  Its host
galaxy is lensed into an Einstein ring (Peng et al.\
\citeyear{pen06}).  It has a supermassive black hole mass of $M_{\rm
BH}$=2.4$\times$10$^9$\,\msol, and follows the relation between black
hole mass and rest-frame $R$-band luminosity of the host galaxy
($M_{\rm BH}$--$L_{\rm R}$) for $z$$\gtrsim$1.7 galaxies (Peng et al.\
\citeyear{pen06}).  Estimating its bolometric AGN luminosity from its
$M_{\rm BH}$ and $B$-band magnitude (Peng et al.\ \citeyear{pen06}),
HE\,1104-1805 follows the $L_{\rm bol}$--$L_{\rm FIR}$ relation for PG
quasars, and is consistent with the values found for other high-$z$
FIR-luminous quasars within the uncertainties (e.g., Omont et al.\
\citeyear{omo03}; Wang et al.\ \citeyear{wan08}; see Tab.~\ref{t3} for
$L_{\rm FIR}$).  HE\,1104--1805 is radio-quiet, but has a luminous
dust bump, with $S_{850\,{\mu}{\rm m}}$=14.8$\pm$3.0\,mJy and
$S_{1.3\,{\rm mm}}$=5.3$\pm$0.9\,mJy. It is magnified by a factor of
$\mu_{\rm L}$=10.8 (Barvainis \& Ivison
\citeyear{bi02}). It also exhibits luminous PAH emission (Lutz et al.\
\citeyear{lut08}).

\subsubsection{New Results}

We detect luminous, spatially resolved \cco\ emission toward both lens
images of HE\,1104--1805 at $>$8$\sigma$ and $>$4$\sigma$ significance
(Fig.~\ref{f1}). From Gaussian fitting to the line profile, we find
$S_\nu$=16.1$\pm$2.4\,mJy and $\Delta$$v_{\rm FWHM}$=441$\pm$81\,\kms,
corresponding to $I_{\rm CO}$=7.5$\pm$1.2\,Jy\,\kms\ (see
Tab.~\ref{t2}). We do not detect the underlying continuum down to a
2$\sigma$ limit of $<$1.5\,mJy. The CO line emission yields a host
galaxy redshift of $z_{\rm CO}$=2.3221$\pm$0.0004, which is close to
the optical redshift of the quasar (d$z$=$\|$$z_{\rm CO}$--$z_{\rm
opt}$$\|$$\simeq$0.003).

Assuming a 2\% correction for subthermal excitation of the \cco\ line,
we derive $L'_{\rm CO}$=2.1$\times$10$^{10}$\,($\mu_{\rm
L}$/14.5)$^{-1}$\,K\,\kms\,pc$^2$, and $M_{\rm
gas}$=1.7$\times$10$^{10}$\,($\mu_{\rm L}$/10.8)$^{-1}$\,($\alpha_{\rm
CO}$/0.8)\,\msol\ (see Tab.~\ref{t3}).

\subsubsection{Origin of the CO Emission}

In Figure \ref{f4}, overlays of the \cco\ emission on top of
observed-frame 555\,nm, 814\,nm, and 1.6\,$\mu$m continuum emission in
HE\,1104--1805 are shown ({\em HST} images from Remy et al.\
\citeyear{rem98}; L\'ehar et al.\ \citeyear{leh00}). The CO emission
is clearly associated with the two (pointlike) images of the AGN. The
emission peaks are slightly spatially offset. This may indicate that
the AGN does not reside in the center of the molecular gas
reservoir. However, this small offset may as well be due to
smaller-scale dynamics of the molecular gas that are averaged over in
the CO emission line map. The CO images are consistent with being
resolved along the Einstein radius, but observations at higher spatial
resolution and sensitivity are desirable to study the structure and
dynamics of the lensed gas reservoir in more detail.

\begin{figure*}
\epsscale{1.15}
\plotone{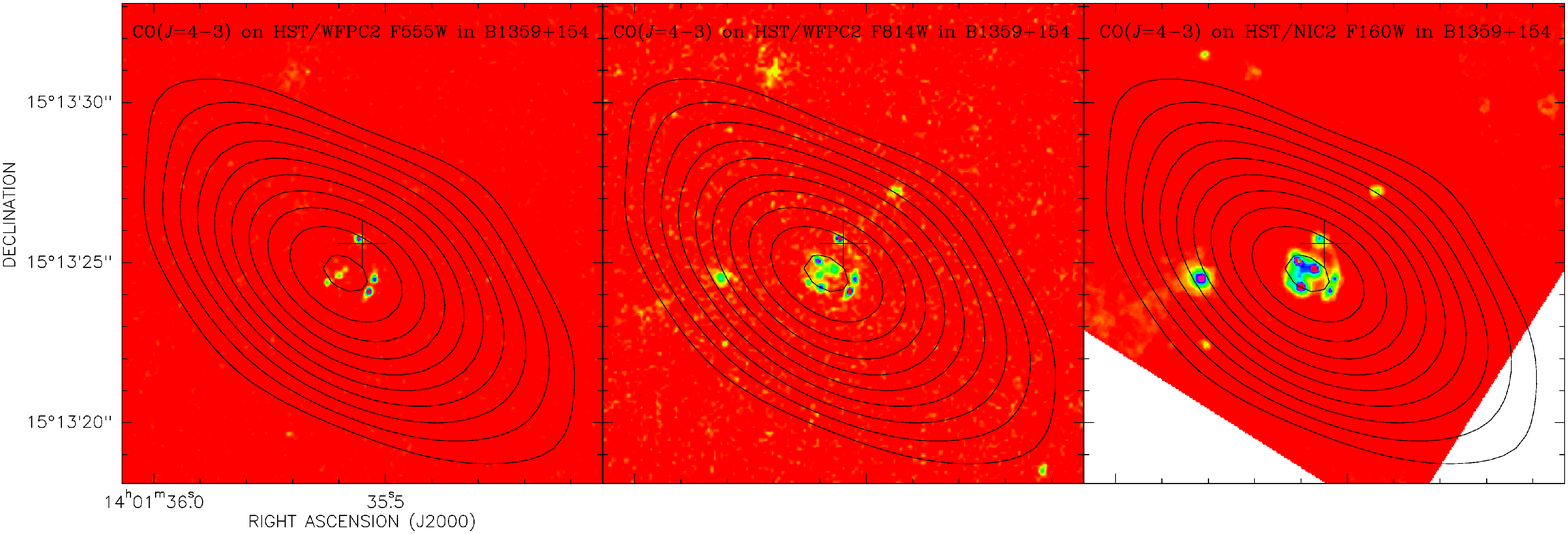}

\caption{Overlays of \dco\ (contours) and 555\,nm, 814\,nm, and 1.6\,$\mu$m 
(rest-frame 131, 192, and 377\,nm) continuum emission ({\em HST}/WFPC2
and NICMOS2, {\em left} to {\em right}; images not cleaned) in
B1359+154. The cross indicates the same position as in Fig.~\ref{f1}.
The six spots visible in all panels are the lens images of the
background quasar, and the three additional spots in the central
region that become visible at longer wavelengths are the foreground
lensing galaxies.
\label{f5}}
%
\end{figure*}

\subsection{B1359+154}

\subsubsection{Previous Results}

B1359+154 is a rare, sextuply lensed, radio-loud quasar at an optical
redshift of \\ $z_{\rm opt}$=3.235$\pm$0.002, with a maximum image
separation of 1.71$''$ (Myers et al.\ \citeyear{mye99}). It was
selected in the radio, but in contrast to B1938+666, is an optically
bright quasar. It is lensed by a compact group at $z_{\rm G}$$\simeq$1
with three primary lensing galaxies of similar luminosities, which are
situated on the vertices of a triangle, separated by $\sim$0.7$''$
(photometric/lensing redshifts are $z_{\rm G1}$=1.35$\pm$0.16, $z_{\rm
G2}$=0.88$\pm$0.06, and $z_{\rm G3}$=0.94$\pm$0.07; Rusin et al.\
\citeyear{rus01}). Very compact (3--9\,mas lensed sizes along their major
axes), flat spectrum ($\alpha_{1.7}^5$$\simeq$--0.3 from 1.7 to
5\,GHz) radio cores are associated with all six images, and the
brightest three images show radio jets extending out to tens of mas
(lensed) scales (Rusin et al.\ \citeyear{rus01}). B1359+154 has a
luminous dust bump, with $S_{450\,{\mu}{\rm m}}$=39$\pm$10\,mJy and
$S_{850\,{\mu}{\rm m}}$=11.5$\pm$1.9\,mJy. It is magnified by a factor
of $\mu_{\rm L}$=118 (Barvainis \& Ivison \citeyear{bi02}).

\subsubsection{New Results:\ Line Emission}

We detect luminous \cco\ and \dco\ emission toward B1359+154
(Fig.~\ref{f1}). From Gaussian fitting to the line profiles, we find
$S_\nu$=5.6$\pm$1.7 and 10.0$\pm$1.6\,mJy, and $\Delta$$v_{\rm
FWHM}$=198$\pm$92 and 237$\pm$47\,\kms\ for the \cco\ and \dco\ lines,
respectively. The line widths are consistent within the errors.  This
corresponds to $I_{\rm CO}$=1.2$\pm$0.4 and 2.5$\pm$0.4\,Jy\,\kms\
(see Tab.~\ref{t2}) . The CO line emission yields a median host galaxy
redshift of $z_{\rm CO}$=3.2399$\pm$0.0003, which is close to the
optical redshift of the quasar (d$z$=$\|$$z_{\rm CO}$--$z_{\rm
opt}$$\|$$\simeq$0.005). The line brightness temperature ratio of
$r_{43}$=1.2$\pm$0.5 is consistent with optically thick, thermalized
emission within the errors.

Assuming a 6\% correction for subthermal excitation of the (higher
signal-to-noise ratio) \dco\ line, we derive $L'_{\rm
CO}$=6.5$\times$10$^8$\,($\mu_{\rm L}$/118)$^{-1}$\,K\,\kms\,pc$^2$,
and \\ 
$M_{\rm gas}$=5.2$\times$10$^8$\,($\mu_{\rm
L}$/118)$^{-1}$\,($\alpha_{\rm CO}$/0.8)\,\msol\ (see Tab.~\ref{t3}),
comparable to what is found in the lensed Ly-break galaxy
MS\,1512-cB58 (Riechers et al.\ \citeyear{rie10b}). Under the
assumption of $\mu_{\rm L}$=20, $L'_{\rm CO}$ and $M_{\rm gas}$ would
be $\lesssim$6$\times$ higher, but still at the low end of observed
values at high $z$.

\subsubsection{New Results:\ Continuum Emission}

As shown in Fig.~\ref{f1}, we detect continuum emission toward
B1359+154 under the CO lines. Simultaneous fits to the line and
continuum emission suggest continuum strengths of 3.97$\pm$0.35 and
2.11$\pm$0.32\,mJy below the \cco\ and \dco\ lines. The strength and
slope of the emission is consistent with synchrotron emission from the
radio-loud AGN. By averaging the line-free channels over the observed
frequency range (LSB+USB for both lines) and fitting a 2-dimensional
Gaussian to the $u-v$ data, we find a 3.1\,mm continuum flux of
3.61$\pm$0.18\,mJy.  In Fig.~\ref{f2}, a map of the averaged continuum
emission is shown.  From the fit, we also find a continuum size of
(3.14$''$$\pm$2.48$''$)$\times$(0.88$''$$\pm$0.72$''$) at a position
angle of 289$^\circ$$\pm$19$^\circ$, suggesting that the continuum
emission is marginally resolved at best. Accounting for
interferometric seeing (which is not part of the above errors), this
is consistent with the 1.7$''$ maximum image separation, and the
continuum emission to be dominated by the radio-loud AGN.

\subsubsection{Origin of the CO/Continuum Emission}

In Figure \ref{f5}, overlays of the \dco\ and underlying continuum
emission on top of observed-frame 555\,nm, 814\,nm, and 1.6\,$\mu$m
continuum emission in B1359+154 are shown ({\em HST} images from Munoz
et al.\ \citeyear{mun98}). The peak of the CO/continuum emission is
consistent with the centroid of the six lensed images of the
source. Observations at (at least) 5$\times$ higher resolution are
desirable to study the distribution and dynamics of the molecular gas
in this complex system.

\section{Discussion}

\begin{figure*}
\epsscale{1.0}
\plotone{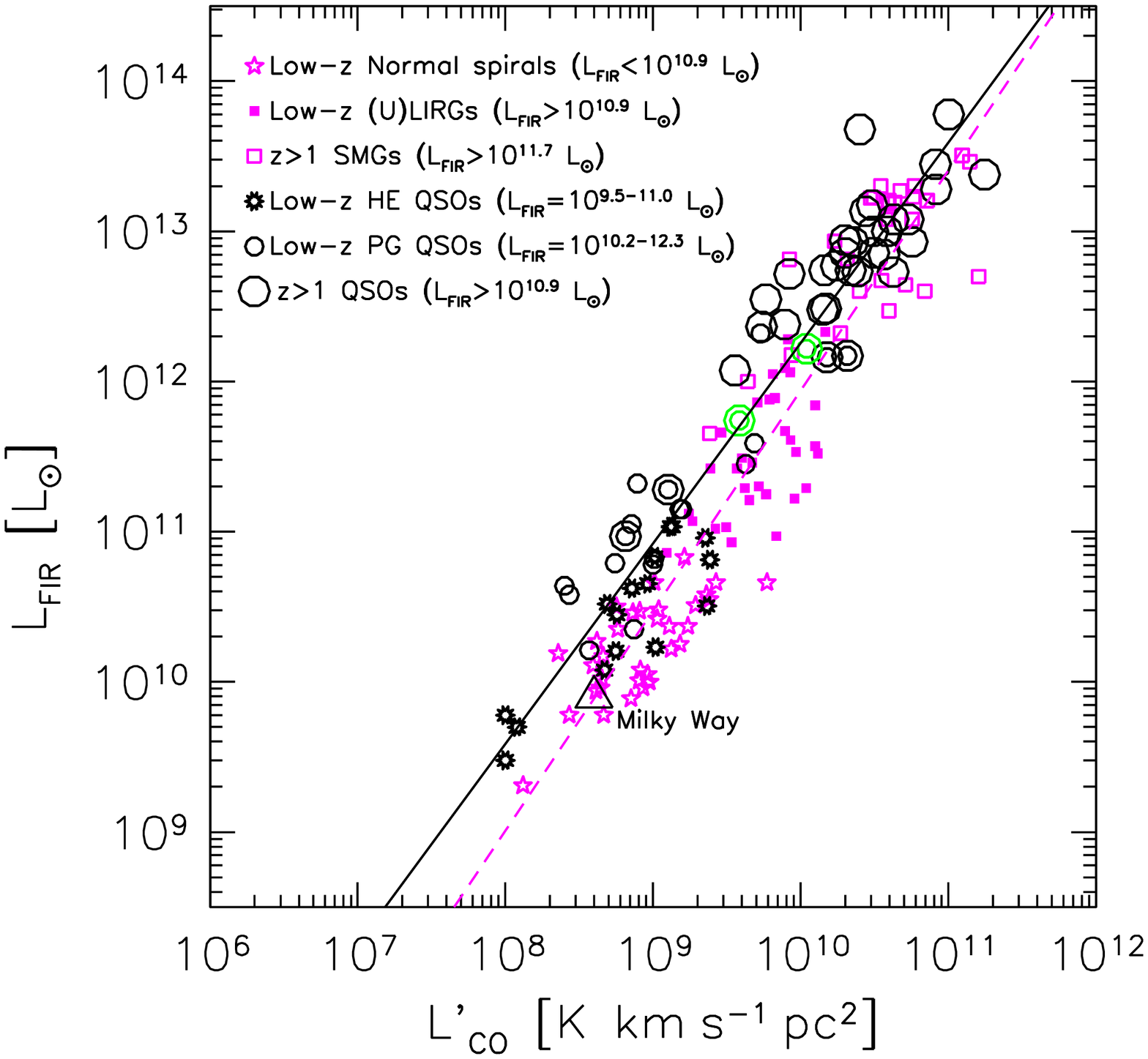}

\caption{Comparison of CO line luminosity with FIR luminosity for samples of 
low-$z$ spiral and starburst galaxies and (U)LIRGs (Gao \& Solomon
\citeyear{gs04}), low-$z$ HE and PG QSOs (Evans et al.\
\citeyear{eva01}, \citeyear{eva06}; Scoville et al.\ \citeyear{sco03};
Bertram et al.\ \citeyear{ber07}), and high-$z$ SMGs (Solomon \&
Vanden Bout \citeyear{sv05}; Tacconi et al.\ \citeyear{tac06}; Iono et
al.\ \citeyear{ion06}; Frayer et al.\ \citeyear{fra08},
\citeyear{fra10}; Schinnerer et al.\ \citeyear{sch08}; Daddi et al.\
\citeyear{dad09a}, \citeyear{dad09b}; Knudsen et al.\
\citeyear{knu09}; Wei\ss\ et al.\ \citeyear{wei09}; Bothwell et al.\
\citeyear{bot10}; Carilli et al.\ \citeyear{car10}; Riechers et al.\
\citeyear{rie10}; Harris et al.\ \citeyear{har10}) and QSOs (Solomon
\& Vanden Bout \citeyear{sv05}; Riechers et al.\ \citeyear{rie06},
\citeyear{rie09a}; Carilli et al.\ \citeyear{car07}; Maiolino et al.\
\citeyear{mai07}; Willott et al.\ \citeyear{wil07}; Aravena et al.\
\citeyear{ara08}; Coppin et al.\ \citeyear{cop08}; Wang et al.\
\citeyear{wan10}). The new lensed high-$z$ QSOs in this paper are
shown as double circles. For comparison, the green circles assume
$\mu_{\rm L}$=20 for B1359+154 and B1938+666. All data are corrected
for gravitational lensing. The solid line is a fit to all quasar
samples (black symbols), corresponding to log\,$L_{\rm FIR}$ =
(1.33$\pm$0.05) log\,$L'_{\rm CO}$--(1.09$\pm$0.50). The dashed line
is a fit to all sources without a dominant AGN (magenta symbols),
corresponding to log\,$L_{\rm FIR}$ = (1.47$\pm$0.06) log\,$L'_{\rm
CO}$--(2.73$\pm$0.54). A fit to all data yields $L_{\rm FIR}$ =
(1.40$\pm$0.04) log\,$L'_{\rm CO}$--(1.92$\pm$0.42).
\label{f7}}
%
\end{figure*}

\subsection{The `Star Formation Law' for AGN Host Galaxies}

Due to high lensing magnification factors, our targets are among the
intrinsically faintest high--$z$ sources observed to date. Their
$L_{\rm FIR}$/$L'_{\rm CO}$ ratios lie at the low end of the
distribution of high-$z$ quasars, but are consistent with the observed
range.

\begin{figure*}
\epsscale{1.15}
\plotone{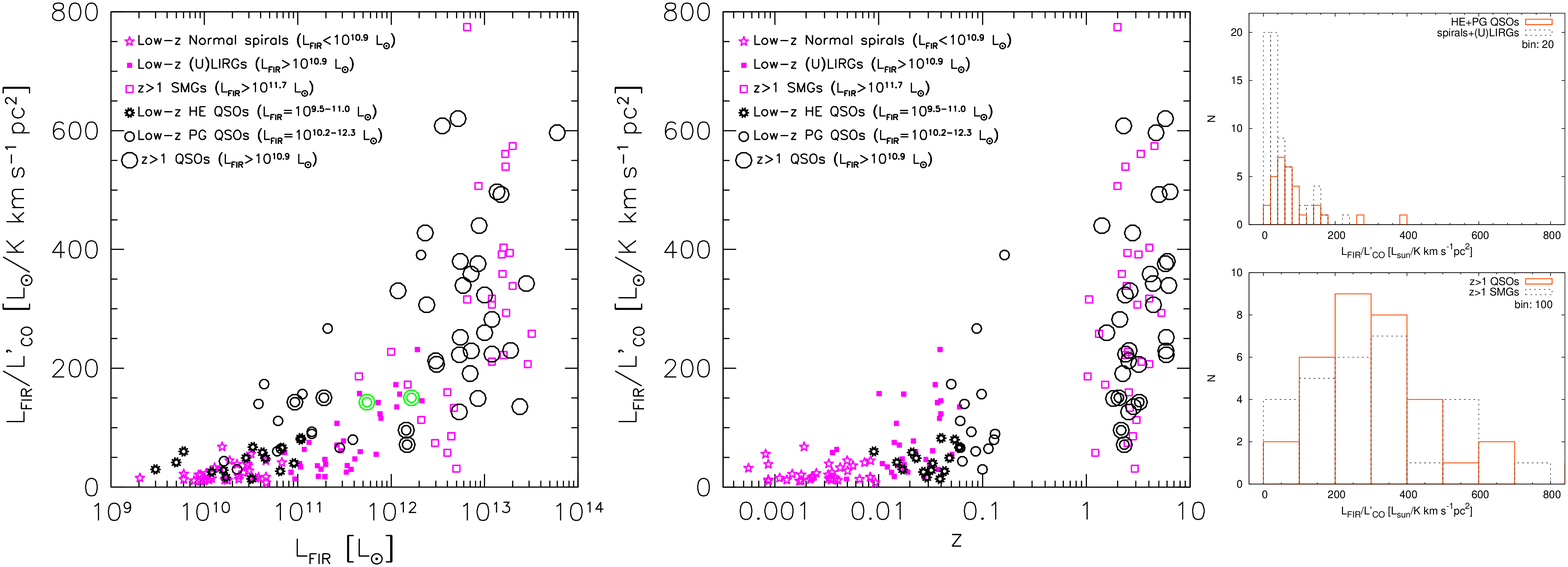}

\caption{$L_{\rm FIR}$/$L'_{\rm CO}$, a measure of star formation 
efficiency, as a function of $L_{\rm FIR}$ ({\em left}) and redshift
({\em middle}), and as histograms for the low-$z$ and high-$z$ samples
({\em right}). The ratio is shown on a linear scale to emphasize the
scatter of the $L'_{\rm CO}$--$L_{\rm FIR}$ relation. The same samples
are shown as in Fig.~\ref{f7}.
\label{f8}}
%
\end{figure*}

To investigate the properties of our targets in more detail, we here
compare their $L'_{\rm CO}$ and $L_{\rm FIR}$ to other CO-detected
quasars at low and high redshift, and low and high redshift galaxies
without luminous AGN see Fig.~\ref{f7}). The $L'_{\rm CO}$--$L_{\rm
FIR}$ relation is a variant of the spatially integrated `star
formation law' (e.g., Kennicutt \citeyear{ken98a}; \citeyear{ken98b})
that only relies on observed quantities, as it is independent of
quantities like the conversion factor $\alpha_{\rm CO}$ and the
assumed stellar initial mass function (see, e.g., Daddi et al.\
\citeyear{dad10}; Genzel et al.\ \citeyear{gen10}, for recent
investigations of the $M_{\rm gas}$--SFR relation). At high redshift,
we restrict the analysis to a comparison of quasars (summarized in
Table~\ref{t4}) and submillimeter galaxies (SMGs), as these samples
cover a comparable range in $L_{\rm FIR}$.\footnote{The high-$z$
massive, gas-rich star-forming galaxies identified by Daddi et al.\
(\citeyear{dad08}, \citeyear{dad10b}) and Tacconi et al.\
(\citeyear{tac10}) have comparable $L'_{\rm CO}$, but substantially
lower $L_{\rm FIR}$/$L'_{\rm CO}$, than these more extreme systems.}
Many SMGs are known to host AGN, but these AGN are substantially less
luminous than those found in quasars.  At low $z$, the galaxies
without luminous AGN are represented by a revised version of the Gao
\& Solomon (\citeyear{gs04}) sample, including luminous and
ultra-luminous infrared galaxies ((U)LIRGs). This currently is the
best-studied, best calibrated low-$z$ galaxy sample for the spatially
integrated galaxy properties studied here. We corrected all high-$J$
CO line-based luminosities for excitation, based on models for
high-$z$ quasars with well-sampled CO line ladders (Riechers et al.\
\citeyear{rie06}; \citeyear{rie09b}), corresponding to corrections of
0\%, 2\%, 6\%, 12\%, and 25\% for the CO $J$=2$\to$1, 3$\to$2,
4$\to$3, 5$\to$4, and 6$\to$5 lines. A fit to all data yields $L_{\rm
FIR}$ = (1.40$\pm$0.04) log\,$L'_{\rm CO}$--(1.92$\pm$0.42), yielding
the canonical power law slope of 1.4 (e.g., Kennicutt
\citeyear{ken98a}; see also Riechers et al.\ \citeyear{rie06}).
Fitting all quasar samples yields log\,$L_{\rm FIR}$ = (1.33$\pm$0.05)
log\,$L'_{\rm CO}$--(1.09$\pm$0.50), i.e., a slightly lower power law
slope. A fit to the low-$z$ galaxy sample and the SMGs yields
log\,$L_{\rm FIR}$ = (1.47$\pm$0.06) log\,$L'_{\rm
CO}$--(2.73$\pm$0.54), i.e., a slightly higher power law slope.

The high-$z$ quasars and SMG samples show comparable $L_{\rm
FIR}$/$L'_{\rm CO}$ ratios and span a comparable range (see also
Fig.~\ref{f8}, where the ratio of both quantitites is shown on a
linear scale to emphasize the scatter of individual sources). On
average, SMGs have somewhat higher ratios (by $\sim$28\% at $L_{\rm
FIR}$=10$^{13}$\,\lsol ), which is likely due to the fact that the
typically \cco\-based $L'_{\rm CO}$ are under-corrected with the 2\%
correction factor for quasars. Recent studies of \aco\ and \bco\
emission in SMGs suggest that a correction factor of 30\%--40\% may be
more appropriate for a substantial fraction of SMGs (e.g., Ivison et
al.\ \citeyear{ivi10}; Carilli et al.\ \citeyear{car10}; Riechers et
al.\ \citeyear{rie10}; Harris et al.\ \citeyear{har10}; Frayer et al.\
\citeyear{fra10}), which would make the $L_{\rm FIR}$/$L'_{\rm CO}$
ratios for quasars and SMGs even more similar on average. However,
studies of the CO excitation in SMGs also show that the emission seen
in low-$J$ lines may be partially associated with a low-excitation gas
component, i.e., material that likely is not concentrated in the
regions where the massive starbursts take place (Carilli et al.\
\citeyear{car10}; Riechers et al.\
\citeyear{rie10}). This is not the case in quasars, which can typically 
be modeled with a single, highly excited gas component down to the
low-$J$ CO lines (e.g., Riechers et al.\ \citeyear{rie06}; Wei\ss\ et
al.\ \citeyear{wei07}). Studies of \aco\ emission in larger samples of
high-$z$ quasars and SMGs are required to investigate these
differences, which directly affect the physical interpretation of the
$L'_{\rm CO}$--$L_{\rm FIR}$ relation. In fact, observations of CO
line ladders down to \aco\ may be a powerful tool to distinguish
different high-$z$ galaxy populations based on their gas properties
alone (e.g., Dannerbauer et al.\ \citeyear{dan09}; Aravena et al.\
\citeyear{ara10}; Riechers et al.\ \citeyear{rie10b}).

Another important aspect is that we here study $L_{\rm FIR}$ (i.e.,
rest-frame 42.5--122.5\,$\mu$m emission), not $L_{\rm IR}$
(8--1000\,$\mu$m). This has a relatively small impact for galaxies
without luminous AGN, but it makes substantial difference for
quasars. Quasars typically show both a warm dust bump with
temperatures of $T_{\rm dust}$=30--60\,K (i.e., comparable to ULIRGs
and SMGs), and a hot dust bump with $T_{\rm dust}$$>$100\,K (likely
associated with the AGN, e.g., Wei\ss\ et al.\
\citeyear{wei03}; Beelen et al.\
\citeyear{bee06}). This yields a $L_{\rm IR}$/$L_{\rm FIR}$ ratio of
typically a factor of a few. Thus, a $L'_{\rm CO}$--$L_{\rm IR}$
relation for quasars would likely be biased toward the AGN
properties. On the other hand, our findings for the $L'_{\rm
CO}$--$L_{\rm FIR}$ relation appear to suggest that, indeed, the FIR
luminosity in high-$z$ quasars is dominated by dust-reprocessed
emission from young stars in the host galaxy, rather than the AGN.

One caveat to the present interpretation of the $L'_{\rm CO}$--$L_{\rm
FIR}$ relation at high $z$ is the possibility of differential
gravitational lensing effects between the CO and FIR emission for the
lensed galaxies in both samples. Given the overall agreement between
lensed and unlensed sources within the samples and the fact that they
follow the $L'_{\rm CO}$--$L_{\rm FIR}$ relation, such effects appear
to be minor, but only few lensed systems have been studied at
sufficient spatial resolution in molecular gas {\em and} FIR continum
emission to quanitfy this in more detail.

The difference in $L_{\rm FIR}$/$L'_{\rm CO}$ ratios between low-$z$
quasar samples and low-$z$ galaxies without luminous AGN appear to be
larger than those between high-$z$ quasars and SMGs. With the current
sample sizes, these differences are only moderately statistically
significant. Given the less massive dust and gas reservoirs in the
lower luminosity systems, such a trend may however not be surprising.
In such systems, the AGN may be substantially more efficient in
heating larger fractions of the warm dust than in the typically very
gas-rich high-$z$ systems, yielding elevated $L_{\rm FIR}$/$L'_{\rm
CO}$ ratios. High spatial resolution observations of the CO and FIR
continuum in nearby quasar host galaxies are key to investigate this
issue in more detail. An alternative interpretation would be that the
AGN fraction rises with $L_{\rm FIR}$ toward ULIRGs and SMGs, which
could decrease the difference between FIR-luminous quasars and other
FIR-luminous galaxy populations. We consider this explanation less
compelling, as there is no clear relation between AGN luminosity and
FIR luminosity for galaxies of comparable masses.

\begin{deluxetable*}{ l c c c c c c c l }
\tabletypesize{\scriptsize}
\tablecaption{Molecular Line and FIR continuum properties of high-$z$ quasars. \label{t4}}
\tablehead{
Source        & $z_{\rm CO}$ & $\mu_{\rm L}$ & CO line & $I_{\rm CO}$ & log$_{10}$($L'_{\rm CO}$)\tablenotemark{a}$^{,}$\tablenotemark{b}  & log$_{10}$($L_{\rm FIR}$)\tablenotemark{a}  & ref.\ & comments \\
              &              &               &         &  [Jy\,\kms ] & [K \kms pc$^2$]           & [\lsol ]                  &       & }
\startdata    

Q0957+561       & 1.4141  & 1.6  & 2$\to$1 & 1.2$\pm$0.1     & 10.30 & 12.94 & 1  & double lens \\
3C318           & 1.571   &      & 2$\to$1 & 1.19$\pm$0.22   & 10.59 & 13.00 & 2  & \\
COSBO11         & 1.8275  &      & 2$\to$1 & 1.33            & 10.76 & 12.93 & 3  & \\
B1938+666       & 2.0590  & 173  & 3$\to$2 & 9.1$\pm$1.1     & 9.10  & 11.28 & 4  & Einstein ring \\
HS\,1002+4400   & 2.1015  &      & 3$\to$2 & 1.7$\pm$0.3     & 10.63 & 13.08 & 5  & \\
HE\,0230--2130  & 2.1664  & 14.5 & 3$\to$2 & 8.3$\pm$1.2     & 10.18 & 12.16 & 4  & quadruple lens \\
RX\,J1249--0559 & 2.2470  &      & 3$\to$2 & 1.3$\pm$0.4     & 10.56 & 12.85 & 5  & \\
F10214+4724     & 2.2858  & 17   & 3$\to$2 & 3.40$\pm$0.19   & 9.76  & 12.55 & 6  & arc lens \\
HE\,1104--1805  & 2.3221  & 10.8 & 3$\to$2 & 7.5$\pm$1.2     & 10.32 & 12.17 & 4  & double lens \\
J1543+5359      & 2.3698  &      & 3$\to$2 & 1.0$\pm$0.2     & 10.49 & 13.00 & 5  & \\
HS\,1611+4719   & 2.3961  &      & 3$\to$2 & 1.7$\pm$0.3     & 10.73 & 13.08 & 5  & \\
J13122+12381    & 2.5564  &      & 3$\to$2 & 0.4$\pm$0.1     & 10.15 & 12.48 & 5  & \\
Cloverleaf      & 2.55784 & 11   & 3$\to$2 & 13.2$\pm$0.2    & 10.63 & 12.73 & 7  & quadruple lens \\
J1409+5628      & 2.5832  &      & 3$\to$2 & 2.3$\pm$0.2     & 10.92 & 13.28 & 8  & \\
MG\,0414+0534   & 2.639   & 27   & 3$\to$2 & 2.6             & 9.55  & 12.07 & 9  & quadruple lens \\
RX\,J0911+0551  & 2.796   & 22   & 3$\to$2 & 2.9$\pm$1.1     & 9.73  & 12.37 & 10 & quadruple lens \\
J04135+10277    & 2.846   & 1.3  & 3$\to$2 & 5.4$\pm$1.3     & 11.25 & 13.38 & 10 & cluster lens \\
MG\,0751+2716   & 3.1999  & 16   & 3$\to$2 & 4.6$\pm$0.5     & 10.17 & 12.49 & 11 & quadruple lens \\
B1359+154       & 3.2399  & 118  & 4$\to$3 & 2.5$\pm$0.4     & 8.81  & 10.97 & 4  & sextuple lens \\
APM\,08279+5255 & 3.9118  & 4.2  & 1$\to$0 & 0.168$\pm$0.015 & 10.40 & 13.68 & 12 & triple lens \\
PSS\,J2322+1944 & 4.1192  & 5.3  & 1$\to$0 & 0.155$\pm$0.013 & 10.30 & 12.86 & 13 & Einstein ring \\
BRI\,1335--0417 & 4.4074  &      & 2$\to$1 & 0.43$\pm$0.02   & 10.91 & 13.45 & 14 & \\
BRI\,0952--0115 & 4.4337  & 4    & 5$\to$4 & 0.91$\pm$0.11   & 9.89  & 12.38 & 15 & double lens \\
BR\,1202--0725  & 4.6949  &      & 1$\to$0 & 0.120$\pm$0.010 & 11.00 & 13.78 & 13 & two sources \\
J0338+0021      & 5.0267  &      & 5$\to$4 & 0.73$\pm$0.09   & 10.48 & 13.18 & 16 & \\
J0927+2001      & 5.7722  &      & 5$\to$4 & 0.44$\pm$0.07   & 10.35 & 12.93 & 17 & \\
J1044--0125     & 5.7824  &      & 6$\to$5 & 0.21$\pm$0.04   & 9.92  & 12.72 & 18 & \\
J0840+5624      & 5.8437  &      & 5$\to$4 & 0.60$\pm$0.07   & 10.50 & 12.86 & 18 & \\
J1425+3254      & 5.8918  &      & 6$\to$5 & 0.59$\pm$0.11   & 10.38 & 12.73 & 18 & \\
J1335+3533      & 5.9012  &      & 6$\to$5 & 0.53$\pm$0.07   & 10.34 & 12.74 & 18 & \\
J2054--0005     & 6.0379  &      & 6$\to$5 & 0.34$\pm$0.07   & 10.16 & 12.74 & 18 & \\
J1048+4637      & 6.2274  &      & 6$\to$5 & 0.39$\pm$0.08   & 10.24 & 12.77 & 18 & \\
J1148+5251      & 6.4189  &      & 3$\to$2 & 0.20$\pm$0.02   & 10.48 & 13.13 & 19 & 
\vspace{-1mm}
\enddata 
\tablerefs{${}$[1] Krips et al.\ \citeyear{kri05}; [2] Willott et al.\ \citeyear{wil07}; [3] Aravena et al.\ \citeyear{ara08}; [4] this work;
[5] Coppin et al.\ \citeyear{cop08}; [6] Ao et al.\ \citeyear{ao08}; [7] Wei\ss\ et al.\ \citeyear{wei03}; [8] Beelen et al.\ \citeyear{bee04};
[9] Barvainis et al.\ \citeyear{bar98}; [10] Hainline et al.\ \citeyear{hai04}; [11] Alloin et al.\ \citeyear{all07}; [12] Riechers et al.\ \citeyear{rie09a};
[13] Riechers et al.\ \citeyear{rie06}; [14] Riechers et al.\ \citeyear{rie08}; [15] Guilloteau et al.\ \citeyear{gui99}; [16] Maiolino et al.\ \citeyear{mai07};
[17] Carilli et al.\ \citeyear{car07}; [18] Wang et al.\ \citeyear{wan10}; [19] Walter et al.\ \citeyear{wal03}.}
\tablecomments{${}$ Estimates are typically based on the lowest-$J$ CO line observed, if applicable.}
\tablenotetext{a}{Corrected for gravitational lensing.}
\tablenotetext{b}{Corrected for line excitation where applicable. Corrections applied:\ 0\% for \bco, 2\% for \cco, 6\% for \dco, 12\% for \eco, and 25\% for \fco.}
\end{deluxetable*}


\subsection{Star Formation Rates and Gas Depletion Timescales}

If the $L_{\rm FIR}$ in our targets are indeed dominated by star
formation, we can derive their star-formation rates (SFRs). Assuming
SFR[\msol\,yr$^{-1}$]=1.5$\times$10$^{-10}$\,$L_{\rm FIR}$[\lsol ]
(e.g., Kennicutt \citeyear{ken98a}; \citeyear{ken98b}), we find SFRs
of 30\,($\mu_{\rm L}$/173)$^{-1}$\,\msol\,yr$^{-1}$, 220\,($\mu_{\rm
L}$/14.5)$^{-1}$\,\msol\,yr$^{-1}$, 220\,($\mu_{\rm
L}$/10.8)$^{-1}$\,\msol\,yr$^{-1}$, and 14\,($\mu_{\rm
L}$/118)$^{-1}$\,\msol\,yr$^{-1}$ for B1938+666, HE\,0230--2130,
HE\,1104--1805, and B1359+154, respectively (Tab.~\ref{t3}).

The minimum times for which the starbursts can be maintained at their
current rates are given by the gas depletion timescales. We find
$\tau_{\rm dep}$=$M_{\rm gas}$/SFR=35, 55, 75, and 40\,Myr for
B1938+666, HE\,0230--2130, HE\,1104--1805, and B1359+154, respectively.
This is comparable to what is found in SMGs and other high-$z$ QSOs
(e.g., Greve et al.\ \citeyear{gre05}; Riechers et al.\
\citeyear{rie08}; \citeyear{rie10}).

\subsection{Systemic Redshifts}

We find that the CO redshifts of the three sources with spectroscopic
redshifts from optical observations are close to $z_{\rm
opt}$. However, in all cases, the CO emission is redshifted relative
to broad Ly$\alpha$ and high ionization (typically C{\scriptsize
IV}$\lambda$1549) optical emission lines from the AGN. The median
redshift difference of the three sources with $z_{\rm opt}$ is
d$z$=0.0045$\pm$0.0015, or 360$\pm$116\,\kms.\footnote{For
consistency, we here adopt $z_{\rm Ly\alpha}$=2.3172 for
HE\,1104--1805 as measured by Smette et al.\ (\citeyear{sme95})
instead of the H$\beta$+[O{\scriptsize III}]($\lambda$5007) redshift
used above.} For a large sample of Sloan Digital Sky Survey (SDSS)
quasars, Richards et al.\ (\citeyear{ric02}) have found that
low-ionization lines such as Mg{\scriptsize II}($\lambda$2798) are
commonly redshifted relative to high-ionization broad lines such as
C{\scriptsize IV}($\lambda$1549) (median of 824$\pm$511\,\kms ), which
they interpret to be due to orientation effects related to the AGN
accretion disk (or disk wind). Even though the orientation of the
accretion disk and host galaxy need not be aligned, referencing the
redshifts of broad AGN lines to the systemic redshifts provided by CO
observations of the host galaxies may help to better disentangle
orientation effects and dynamics of high redshift galaxies.

\section{Summary}

We have detected \bco, \cco, and \dco\ emission in four lensed quasar
host galaxies at redshifts of $z$$>$2 with CARMA (a total of six
lines). To date, these are the highest redshift CO detections reported
with CARMA. From our analysis, we obtain the following key results:\

Facilitating the large, $>$7.4\,GHz bandwidth (LSB+USB) of the new
CARMA correlator, we have executed the first successful `blind' CO
redshift search with an interferometer, demonstrating the feasibility
of such investigations with only few frequency setups. Based on the CO
redshift, we also used the broad frequency range covered by these
observations to search for emission from dense molecular gas tracers
(HCN, HCO$^+$, HNC, C$_2$H, and CN), which yielded only a statistical
detection after stacking -- as expected at the depth of this
search. We also searched for HCN, HCO$^+$, and HNC absorption toward
foreground sources, but no lines are detected along the line of sight.

We spatially resolve the CO emission in the two radio-quiet quasars in
the sample. In the wide separation double lens HE\,1104--1805
($z$=2.322), we individually detect the emission toward both
images. In all cases the CO emission is consistent with the brightest
emission regions detected at rest-frame optical wavelengths.

We detect strong continuum emission toward the two radio-loud sources
in the sample. The spectral slope of the emission is consistent with
synchrotron emission associated with the AGN in both cases.

We derive lensing- and excitation-corrected CO line luminosities of
0.65--21$\times$10$^{9}$\,K\,\kms\,pc$^2$. From the CO luminosities,
we determine molecular gas masses of
0.52--17$\times$10$^{9}$\,\msol. These values are at the lower end of
those observed in other high-$z$ quasars, showing the advantage of
observing lensed distant galaxies to probe down to lower gas masses.

Combining our targets with literature samples, we find no significant
difference in the $L_{\rm FIR}$/$L'_{\rm CO}$ ratios between high-$z$
quasars and SMGs. We find tentative evidence that nearby quasars with
low $L_{\rm FIR}$ show an excess in $L_{\rm FIR}$/$L'_{\rm CO}$
relative to systems without a luminous AGN at similar redshift and
with comparable $L'_{\rm CO}$. This may indicate that AGN heating of
the warm dust is more efficient than in the more gas-rich and massive
FIR-luminous quasars detected at high $z$, where the dust heating in
the FIR appears to be largely dominated by young stars.\\[-3mm]

The observations presented herein lay the foundation for future
studies with the Atacama Large (sub-) Millimeter Array (ALMA), which
will routinely observe unlensed galaxies at comparable and fainter
intrinsic CO luminosities. The `blind' CO redshift search technique
presented here will be particularly valuable for so-called `Deep
Field' investigations with ALMA, which will allow, for the first time,
to construct an unbiased CO luminosity function out to high $z$ by
directly selecting galaxies through their CO content. This is an
important step toward constraining the gas mass history of the
universe, a critical piece in our understanding of galaxy evolution
throughout cosmic times.

\acknowledgments 
We thank the referee for a helpful and constructive report. DR
acknowledges support from from NASA through Hubble Fellowship grant
HST-HF-51235.01 awarded by the Space Telescope Science Institute,
which is operated by the Association of Universities for Research in
Astronomy, Inc., for NASA, under contract NAS 5-26555. Support for
CARMA construction was derived from the G.\ and B.\ Moore Foundation,
the K.~T.\ and E.~L.\ Norris Foundation, the Associates of the
California Institute of Technology, the states of California,
Illinois, and Maryland, and the NSF. Ongoing CARMA development and
operations are supported by the NSF under a cooperative agreement, and
by the CARMA partner universities.


\end{document}